\documentclass[aps,prd,superscriptaddress,11pt,notitlepage,nofootinbib]{revtex4-1} 

\usepackage[left=1in,right=1in]{geometry}
\usepackage[utf8]{inputenc}
\usepackage{graphicx}
\usepackage{amsmath,amssymb}
\usepackage{booktabs}
\usepackage{multirow}
\usepackage{physics}
\usepackage{fontawesome}
\usepackage{color}
\definecolor{nicered}{rgb}{0.5,0.,0.}
\definecolor{nicegreen}{rgb}{0.,0.5,0.}
\definecolor{niceblue}{rgb}{0.,0.,0.5}
\usepackage[colorlinks,citecolor=nicegreen,linkcolor=nicered,urlcolor=blue]{hyperref}

\begin{document}
%%%%%%%%%%%%%%%%%%%%%%%%%%%%%%%%%%%%%%%%%%%%
\title{Implications for Electric Dipole Moments of a\\ Leptoquark Scenario for the $B$-Physics Anomalies}
%%%%%%%%%%%%%%%%%%%%%%%%%%%%%%%%%%%%%%%%%%%%
\def\ucsc{University of California, Santa Cruz, and Santa Cruz Institute for Particle Physics, Santa Cruz, CA 95064, USA}
\def\fermilab{Fermi National Accelerator Laboratory, Batavia, IL, 60510, USA}

\author{Wolfgang~Altmannshofer} 
\email[Electronic address: ]{ waltmann@ucsc.edu}
\affiliation{\ucsc}
\author{Stefania~Gori} 
\email[Electronic address: ]{ sgori@ucsc.edu}
\affiliation{\ucsc}
\author{Hiren~H.~Patel} 
\email[Electronic address: ]{ hpatel6@ucsc.edu}
\affiliation{\ucsc}
\author{Stefano~Profumo} 
\email[Electronic address: ]{ profumo@ucsc.edu}
\affiliation{\ucsc}
\author{Douglas~Tuckler} 
\email[Electronic address: ]{ dtuckler@ucsc.edu}
\affiliation{\ucsc}\affiliation{\fermilab}

%%%%%%%%%%%%%%%%%%%%%%%%%%%%%%%%%%%%%%%%%%%%
\begin{abstract}
  Vector leptoquarks can address the lepton flavor universality anomalies in decays associated with the $b \to c \ell \nu$ and $b \to s \ell \ell$ transitions, as observed in recent years. Generically, these leptoquarks yield new sources of CP violation. In this paper, we explore constraints and discovery potential for electric dipole moments (EDMs) in leptonic and hadronic systems. 
  We provide the most generic expressions for dipole moments induced by vector leptoquarks at one loop.  We find that $O(1)$ CP-violating phases in tau and muon couplings can lead to corresponding EDMs within reach of next-generation EDM experiments, and that existing bounds on the electron EDM already put stringent constraints on CP-violating electron couplings.
\end{abstract}
%%%%%%%%%%%%%%%%%%%%%%%%%%%%%%%%%%%%%%%%%%%%

\date{\today}

\maketitle

%%%%%%%%%%%%%%%%%%%%%%%%%%%%%%%%%%%%%%%%%%%%
\section{Introduction} \label{sec:intro}
%%%%%%%%%%%%%%%%%%%%%%%%%%%%%%%%%%%%%%%%%%%%

Over the past several years, multiple $B$-physics experiments, including BaBar, LHCb, and Belle, have reported anomalies in decays associated with the $b \to c \ell \nu$ and $b \to s \ell \ell$ transitions. Violations of lepton flavor universality (LFU), known to be theoretically clean probes of New Physics (NP), are of particular interest. In the Standard Model (SM) LFU is only broken by the lepton masses. Hints for additional sources of LFU violation have been observed in the ratios of branching ratios of flavor-changing charged current and neutral current decays of $B$ mesons, $R_D$, $R_{D^*}$, $R_{K}$, and $R_{K^*}$,
\begin{equation}
    R_{D^{(*)}} = \frac{\text{BR}(B \to D^{(*)} \tau \nu)}{\text{BR}(B \to D^{(*)} \ell \nu)}\,, \qquad R_{K^{(*)}} = \frac{\text{BR}(B \to K^{(*)} \mu^+\mu^-)}{\text{BR}(B \to K^{(*)} e^+e^-)} \,.
\end{equation}
The experimental world averages of $R_D$ and $R_{D^*}$ from the heavy flavor averaging group (HFLAV) are based on measurements from BaBar~\cite{Lees:2012xj}, Belle~\cite{Huschle:2015rga,Hirose:2016wfn,Belle:2019rba}, and LHCb~\cite{Aaij:2015yra,Aaij:2017uff}, and read~\cite{Amhis:2019ckw}
\begin{equation}
    R_D = 0.340 \pm 0.027 \pm 0.013 \,, \qquad R_{D^*} = 0.295 \pm 0.011 \pm 0.008 \,,
\end{equation}
with an error correlation of $\rho = -38\%$.
The corresponding SM predictions are known with high precision~\cite{Bernlochner:2017jka,Bigi:2017jbd,Jaiswal:2017rve}. The values adopted by HFLAV are~\cite{Amhis:2019ckw}
\begin{equation}
    R_D^\text{SM} = 0.299 \pm 0.003  \,, \qquad R_{D^*}^\text{SM} = 0.258  \pm 0.005 \,.
\end{equation}
The combined discrepancy between the SM prediction and experimental world averages of $R_D$ and $R_{D^*}$ is at the $3.1\sigma$ level.

The most precise measurement to date of the LFU ratio $R_K$ has been performed by LHCb~\cite{Aaij:2019wad}
\begin{equation}
 R_K = 0.846^{+0.060}_{-0.054}{}^{+0.016}_{-0.014} \,, \qquad \text{for}~~~ 1.1\,\text{GeV}^2 < q^2 < 6\,\text{GeV}^2 \,,
\end{equation}
with $q^2$ being the dilepton invariant mass squared.
The SM predicts $R_K^\text{SM} \simeq 1$ with theoretical uncertainties well below the current experimental ones~\cite{Bordone:2016gaq}. 
The above experimental value is closer to the SM prediction than the Run-1 result~\cite{Aaij:2014ora}. However, the reduced experimental uncertainties still imply a tension between theory and experiment of $2.5\sigma$.

The most precise measurement of $R_{K^*}$ is from a Run-1 LHCb analysis~\cite{Aaij:2017vbb} that finds
\begin{equation}\label{eq:RKstarEXP}
 R_{K^*} = \begin{cases} 0.66^{+0.11}_{-0.07}\pm0.03 \,, \qquad \text{for}~ 0.045\,\text{GeV}^2 < q^2 < 1.1\,\text{GeV}^2 \,, \\ 0.69^{+0.11}_{-0.07}\pm0.05 \,, \qquad \text{for}~ 1.1\,\text{GeV}^2 < q^2 < 6\,\text{GeV}^2 \,. \end{cases}
\end{equation}
The result for both $q^2$ bins are in tension with the SM prediction~\cite{Bordone:2016gaq}, $R_{K^*}^\text{SM} \simeq 1$, by $\sim 2.5\sigma$ each. Recent measurements of $R_{K^*}$ and $R_K$ by Belle~\cite{Abdesselam:2019wac,Abdesselam:2019lab}\footnote{Here we quote the isospin average of $B^0 \to K^{(*)0} \ell^+ \ell^-$ and $B^\pm \to K^{(*)\pm} \ell^+ \ell^-$ decays.}
\begin{equation}
 R_{K^*} = \begin{cases} 0.90^{+0.27}_{-0.21}\pm0.10 \,, \qquad \text{for}~ 0.1\,\text{GeV}^2 < q^2 < 8\,\text{GeV}^2 \,, \\ 1.18^{+0.52}_{-0.32}\pm0.10 \,, \qquad \text{for}~ 15\,\text{GeV}^2 < q^2 < 19\,\text{GeV}^2 \,, \end{cases}
\end{equation}
\begin{equation}
 R_{K} = \begin{cases} 0.98^{+0.27}_{-0.23}\pm0.06 \,, \qquad \text{for}~ 1\,\text{GeV}^2 < q^2 < 6\,\text{GeV}^2 \,, \\ 1.11^{+0.29}_{-0.26}\pm0.07 \,, \qquad \text{for}~ 14.18\,\text{GeV}^2 < q^2 \,, \end{cases}
\end{equation}
are compatible with both the SM prediction and the LHCb results. Several papers have re-analyzed the status of the $B$ anomalies in light of the latest experimental updates, and found preference for new physics with high significance~\cite{Alguero:2019ptt,Alok:2019ufo,Ciuchini:2019usw,Datta:2019zca,Aebischer:2019mlg,Kowalska:2019ley,Arbey:2019duh}.

While the anomalies detailed upon above persist, the question of the origin of the observed baryon asymmetry \cite{Ade:2015xua} also remains a long standing problem in cosmology.  Any dynamical explanation requires sizable C- and CP-violating interactions in the early universe \cite{Sakharov:1967dj}.  In light of upcoming low-energy experiments with much greater sensitivity to electric and magnetic dipole moments of elementary particles, it is interesting to ask whether solutions to the flavor anomalies may also be associated with sizable CP violating complex phases that may be probed by these experiments.  

The only known viable, single-mediator explanation of all flavor anomalies is a $U_1$ vector leptoquark~\cite{Alonso:2015sja, Bhattacharya:2016mcc, Buttazzo:2017ixm,Calibbi:2017qbu,Crivellin:2018yvo,Angelescu:2018tyl, Cornella:2019hct}.  This leptoquark generically introduces new sources of CP violation in the Lagrangian in the form of complex parameters \cite{Panico:2018hal}.  The scope of the present study is to explore, for the first time, the prospects of observing electric dipole moments (EDMs) induced by a $U_1$ vector leptoquark that could explain the flavor anomalies reviewed above.  We additionally explore collider constraints, as well as constraints from measurements of the magnetic moments, and other flavor observables. Implications for EDMs and other CPV observables in scalar leptoquark scenarios have recently been discussed in~\cite{Bobeth:2017ecx,Fuyuto:2018scm,Dekens:2018bci,Crivellin:2019kps}.

This paper is organized as follows: In Sec.~\ref{sec:model}, we  introduce the CP violating $U_1$ model and discuss its effects on the B-physics anomalies. In Sec.~\ref{sec:dipoles}, we give an overview of the effects of the CP violating leptoquark on EDMs of quarks, leptons, and neutrons. We also include a discussion of the present status of the experimental searches and the prospects for future measurements. In Sec.~\ref{Sec:Pheno}, we report the main results of our paper, showing the leptoquark parameter space that can be probed by B-physics and EDM measurements. In Sec.~\ref{Sec:LHC}, we discuss the LHC bounds on our leptoquark model. Finally, we reserve Sec.~\ref{Sec:Conclusions} for our conclusions.

%%%%%%%%%%%%%%%%%%%%%%%%%%%%%%%%%%%%%%%%%%%%
\section{The CP violating \boldmath $U_1$ Vector Leptoquark Model} \label{sec:model}
%%%%%%%%%%%%%%%%%%%%%%%%%%%%%%%%%%%%%%%%%%%%

We consider the vector leptoquark $U_1 = ({\mathbf 3}, {\mathbf 1})_{2/3}$ (triplet under $SU(3)_c$, singlet under $SU(2)_L$, and with hypercharge $+2/3$).  This model may be viewed as the low energy limit of Pati-Salam models described in Ref.~\cite{Bordone:2017bld, Bordone:2018nbg} (see also~\cite{Assad:2017iib,DiLuzio:2017vat,Barbieri:2017tuq,Blanke:2018sro,Greljo:2018tuh,Heeck:2018ntp,Balaji:2018zna,Fornal:2018dqn}).
The most general dimension-4 Lagrangian describing the vector leptoquark of mass $M_{U_1}$ is (see e.g.~\cite{Dorsner:2016wpm} for a recent review)
\begin{eqnarray} \label{eq:U1}
 \mathcal L_{U_1} &=& -\frac{1}{2} U_{\mu\nu}^\dagger U^{\mu\nu} + M_{U_1}^2 U_\mu^\dagger U^{\mu} \nonumber \\
 && + i g_s U_\mu^\dagger T_a U_\nu \left( \kappa_s G^{\mu\nu}_a + \tilde \kappa_s \tilde G^{\mu\nu}_a \right)+ i g' \frac{2}{3} U_\mu^\dagger U_\nu \left( \kappa_Y B^{\mu\nu} + \tilde \kappa_Y  \tilde B^{\mu\nu}\right) \,, \nonumber \\ 
 && + \sum_{i,j} \left( \lambda_{ij}^q (\bar Q_i \gamma_\mu P_L L_j) U^\mu + \lambda_{ij}^d (\bar D_i \gamma_\mu P_R E_j) U^\mu \right) + \text{h.c.}\,,
\end{eqnarray}
where $U^{\mu\nu} = D^\mu U^\nu - D^\nu U^\mu$ is the leptoquark field strength tensor in terms of its vector potential $U^\mu$ and gauge covariant derivative $D^\mu = \partial^\mu + i g_s T_a G_a^\mu + i g' \frac{2}{3} B^\mu$.   $G_a^\mu$ and $B^\mu$, and $G_a^{\mu\nu}$ and $B^{\mu\nu}$ are the gluon and hypercharge vector potentials and field strengths, respectively.  The dual field strength tensors are $\tilde G_a^{\mu\nu} = \frac{1}{2} \epsilon^{\mu\nu\rho\sigma} G_{a\,\rho\sigma}$ and $\tilde B^{\mu\nu}= \frac{1}{2} \epsilon^{\mu\nu\rho\sigma} B_{\rho\sigma}$.

The third line in Eq.~\eqref{eq:U1} contains couplings of $U_1$ with the SM quarks and leptons. Specifically, $Q_i$ and $L_i$ are the left-handed quark and lepton doublets, while $D_j$ and $E_j$ are the right-handed down quark and charged lepton singlets.
We assume that the model does not contain light right-handed neutrinos. (If right-handed neutrinos are introduced, additional couplings of $U_1$ with right-handed neutrinos and right-handed up quarks are possible~\cite{Robinson:2018gza}.)
The couplings $\lambda_{ij}^q$ and $\lambda_{ij}^d$ are in general complex and are therefore a potential source of CP violation of the model.   We work in the fermion mass eigenstate basis and define the leptoquark couplings $\lambda_{ij}^q$ and $\lambda_{ij}^d$ in a way such that
\begin{equation}\label{eq:yuk}
 \mathcal L_{U_1} \supset \sum_{ijk} (V_{ik} \lambda_{kj}^q) (\bar u_i \gamma_\mu P_L \nu_j) U^\mu +  \sum_{ij} \lambda_{ij}^q (\bar d_i \gamma_\mu P_L \ell_j) U^\mu +  \sum_{ij} \lambda_{ij}^d (\bar d_i \gamma_\mu P_R \ell_j) U^\mu + \text{h.c.}\,,
\end{equation}
where $V$ is the CKM matrix.

The second line in Eq.~\eqref{eq:U1} encodes the chromo- and hypercharge- magnetic and electric dipole moments of the $U_1$ leptoquark.

If the leptoquark arises from the spontaneous breakdown of a gauge symmetry, gauge invariance requires these couplings to be fixed to $\kappa_s = \kappa_Y = 1$, $\tilde \kappa_s = \tilde \kappa_Y = 0$.  In more generic scenarios where $U_1$ is composite, the values of $\kappa_s$, $\tilde \kappa_s$, $\kappa_Y$, $\tilde \kappa_Y$ are free parameters.
Non-zero values for $\tilde \kappa_s$ and $\tilde \kappa_Y$ are an additional potential source of CP violation.  However, since they do not directly influence flavor physics, we will focus our attention to CP-violation contained in $\lambda_{ij}^q$ and $\lambda_{ij}^d$ (even though in Sec.~\ref{sec:dipoles} we will present fully generic expressions for the EDMs, including their dependence on $\tilde \kappa_s$ and $\tilde \kappa_Y$).

 \subsection{Leptoquark Effects in B-meson Decays} \label{sec:LQBdecays}

The $U_1$ leptoquark can simultaneously address the hints for LFU violation in charged current decays $R_{D^{(*)}}$ and in neutral current decays $R_{K^{(*)}}$\footnote{Note that the small anomaly in the low $q^2$ bin of $R_K^*$ in (\ref{eq:RKstarEXP}) cannot be fully addressed by the $U_1$ leptoquark, but it requires the presence of light NP \cite{Sala:2017ihs,Ghosh:2017ber,Datta:2017ezo,Altmannshofer:2017bsz}.}. Here we will use the results of a recent study~\cite{Aebischer:2019mlg} that identified a benchmark point in the leptoquark parameter space that gives a remarkably consistent new physics explanation of these hints. We will explore the parameter space around this benchmark point (supplemented by a few more points), focusing on the implications for dipole moments. As we discuss below, not all leptoquark couplings in (\ref{eq:U1}) are required to address the anomalies.

Explaining the observed values of $R_{D^{(*)}}$ by non-standard effects in the $b \to c \tau \nu$ transition is possible if the leptoquark has sizable couplings to the left-handed tau. Avoiding strong constraints from leptonic tau decays $\tau \to \nu_\tau \ell \bar \nu_\ell$ and the $B \to X_s \gamma$ decay is possible in a well defined parameter space around the benchmark point with $\lambda^q_{33} \simeq 0.7$, $\lambda^q_{23} \simeq 0.6$ with a leptoquark mass of $M_{U_1} = 2$\,TeV~\cite{Aebischer:2019mlg}.
This corresponds to the following non-standard value for $R_{D^{(*)}}$
\begin{equation}
 \frac{R_{D^{(*)}}}{R_{D^{(*)}}^\text{SM}} = \left| 1 + \frac{v^2}{2 M_{U_1}^2} \frac{\lambda^{q*}_{33}\lambda^q_{23}}{V_{cb}}  \right|^2 \simeq 1.2 \,,
\end{equation}
which is in good agreement with observations (in this equation we normalize $v=246$ GeV).

The results for $R_{K^{(*)}}$ can be accommodated by a non-standard effect in the $b \to s \mu\mu$ transition if the couplings to the left-handed muon obey Re$(\lambda^q_{22} \times \lambda^q_{32}) \simeq -2.5 \times 10^{-3}$ for $M_{U_1} = 2$\,TeV~\cite{Aebischer:2019mlg}.
The leptoquark effects for this choice of couplings are described by a shift in the Wilson coefficients of the effective Hamiltonian relevant for $b \to s \ell \ell$ transitions (see e.g.~\cite{Aebischer:2019mlg} for the precise definition)
\begin{equation}
C_9^{bs\mu\mu} = - C_{10}^{bs\mu\mu} = - \frac{4\pi^2}{e^2} \frac{v^2}{M_{U_1}^2} \frac{\lambda_{32}^q \lambda_{22}^{q*}}{V_{ts}^* V_{tb}} \simeq -0.4 \,.
\end{equation}
This agrees well with the best fit value for the Wilson coefficients found in~\cite{Aebischer:2019mlg}.

The muonic couplings $\lambda^q_{22}, \lambda^q_{32}$ (that can explain the $R_{K^{(*)}}$ anomalies) in combination with the tauonic couplings $\lambda^q_{23}, \lambda^q_{33}$ (that are required to explain the $R_{D^{(*)}}$ anomalies) lead to lepton flavor violating decays. The strongest constraints arise from the decays $\tau \to \phi \mu$ and $B \to K \tau \mu$. For the $\lambda_{33}^q,~\lambda_{23}^q$ benchmark mentioned above, existing limits on those decay modes result in the bounds on the leptoquark couplings $|\lambda^q_{22}| \lesssim 0.16$ and $ |\lambda^q_{32}| \lesssim 0.40$ for $M_{U_1} = 2$\,TeV~\cite{Aebischer:2019mlg}.

The experimental values of $R_{K^{(*)}}$ may also be explained by new physics in the $b \to s ee$ transition as opposed to modifying the $b \to s \mu\mu$ transition. Focusing on left-handed couplings, the required shifts in the relevant Wilson coefficients is~\cite{Aebischer:2019mlg}
\begin{equation} \label{eq:C9e}
C_9^{bsee} = - C_{10}^{bsee} = -\frac{4\pi^2}{e^2} \frac{v^2}{M_{U_1}^2} \frac{\lambda_{31}^q \lambda_{21}^{q*}}{V_{ts}^* V_{tb}} \simeq +0.4 \,,
\end{equation}
corresponding to the couplings Re$(\lambda^q_{21} \times \lambda^q_{31}) \simeq +2.5 \times 10^{-3}$ for $M_{U_1} = 2$\,TeV.
The experimental bounds on the lepton flavor violating processes $\tau \to \phi e$ and $B \to K \tau e$ are comparable to those of $\tau \to \phi \mu$ and $B \to K \tau \mu$~\cite{Aubert:2009ap,Miyazaki:2011xe,Lees:2012zz}. We therefore expect that the constraints on the left-handed electron couplings $|\lambda^q_{21}|$ and $|\lambda^q_{31}|$ are similar to the muon couplings mentioned above, i.e. $|\lambda^q_{21}| \lesssim 0.16$ and $ |\lambda^q_{31}| \lesssim 0.40$ for $M_{U_1} = 2$\,TeV.

Motivated by this discussion, in the next sections we will explore the leptoquark parameter space in the neighborhood of four benchmark scenarios:
\begin{subequations}\label{eq:BM}
\begin{align}
\text{BM1:}&~~
\lambda_{33}^q =0.7\,,~\lambda_{23}^q=0.6\,, \quad
\lambda_{32}^q = -0.25\,,~ \lambda_{22}^q = 0.01\,, \quad
\lambda_{31}^q = \lambda_{21}^q = 0\,,\label{eq:BM1} \\ 
\text{BM2:}&~~
\lambda_{33}^q =0.7\,,~\lambda_{23}^q=0.6\,, \quad 
\lambda_{32}^q = \lambda_{22}^q = 0\,, \quad 
\lambda_{31}^q = 0.05\,, \lambda_{21}^q = 0.05\,,\label{eq:BM2}\\ 
\text{BM3:}&~~
\lambda_{33}^q = \lambda_{23}^q=0\,, \quad
\lambda_{32}^q = -1.4\,,~ \lambda_{22}^q = 10^{-3}\,, \quad
\lambda_{31}^q = \lambda_{21}^q = 0\,, \label{eq:BM3}\\ 
\text{BM4:}&~~
\lambda_{33}^q = \lambda_{23}^q=0\,, \quad
\lambda_{32}^q = \lambda_{22}^q = 0 \,, \quad
\lambda_{31}^q = 0.5\,,~ \lambda_{21}^q = 5.0\times 10^{-3}\,, \label{eq:BM4}\\[6pt] 
& ~~M_{U_1} =2~{\rm{TeV}}, \enspace \kappa_{Y,s}=1, \enspace \tilde\kappa_{Y,s}=0 ~\text{for all benchmarks } \nonumber
\end{align}
\end{subequations}
with all the other fermionic couplings of the leptoquark in Eq.~(\ref{eq:U1}) set to zero.
In BM1 and BM2 both the $R_{D^{(*)}}$ and $R_{K^{(*)}}$ anomalies are addressed. The $R_{K^{(*)}}$ explanations involve new physics in the $b \to s \mu\mu$ transition (BM1) or in the $b \to s ee$ transition (BM2). For benchmark points BM3 and BM4 we forgo an explanation of $R_{D^{(*)}}$. This allows us to increase the couplings to muons/electrons while avoiding the strong constraints from lepton flavor violating tau decays. Note that in benchmark BM3, the $R_{K^{(*)}}$ anomalies are only partially addressed. For BM3 we have  $R_K \simeq R_K^{*} \simeq 0.88$ which is in good agreement with the latest $R_K$ measurement, but $\sim 2\sigma$ away from the measured $R_{K^*}$ value. As we discuss below in Sec.~\ref{sec:muon} benchmark BM3 is motivated because it can accommodate the longstanding discrepancy in the anomalous magnetic moment of the muon.

For all benchmark scenarios we explicitly checked compatibility with the measurements of the di-lepton~\cite{Greljo:2017vvb} and di-tau \cite{Faroughy:2016osc} invariant mass distributions at the LHC and searches for electron-quark contact interactions at LEP~\cite{Schael:2013ita}. In the case of the di-lepton invariant mass distributions at the LHC, the value of $\lambda^q_{32}$ in BM3 is close to the exclusion bound.

Starting with these benchmark points, in the following sections we turn on couplings to right-handed taus $\lambda^d_{33}$, muons $\lambda^d_{32}$, and electrons $\lambda^d_{31}$ and determine the expected size of electric and magnetic dipole moments of the leptons as function of the real and imaginary part of the new couplings.  In principle, also the couplings $\lambda^d_{23}$, $\lambda_{22}^d$ and $\lambda_{21}^d$ influence the dipole moments; we comment on $\lambda_{22}^d$ and $\lambda_{21}^d$ in Secs. \ref{sec:muon} and \ref{sec:electron}, but we do not consider $\lambda^d_{23}$ since it does not play any role in explaining the flavor anomalies.
 
%%%%%%%%%%%%%%%%%%%%%%%%%%%%%%%%%%%%%%%%%
\section{Dipole Moments of Quarks and Leptons} \label{sec:dipoles}
%%%%%%%%%%%%%%%%%%%%%%%%%%%%%%%%%%%%%%%%%

\begin{figure}
\centering
\includegraphics[width=0.50\textwidth]{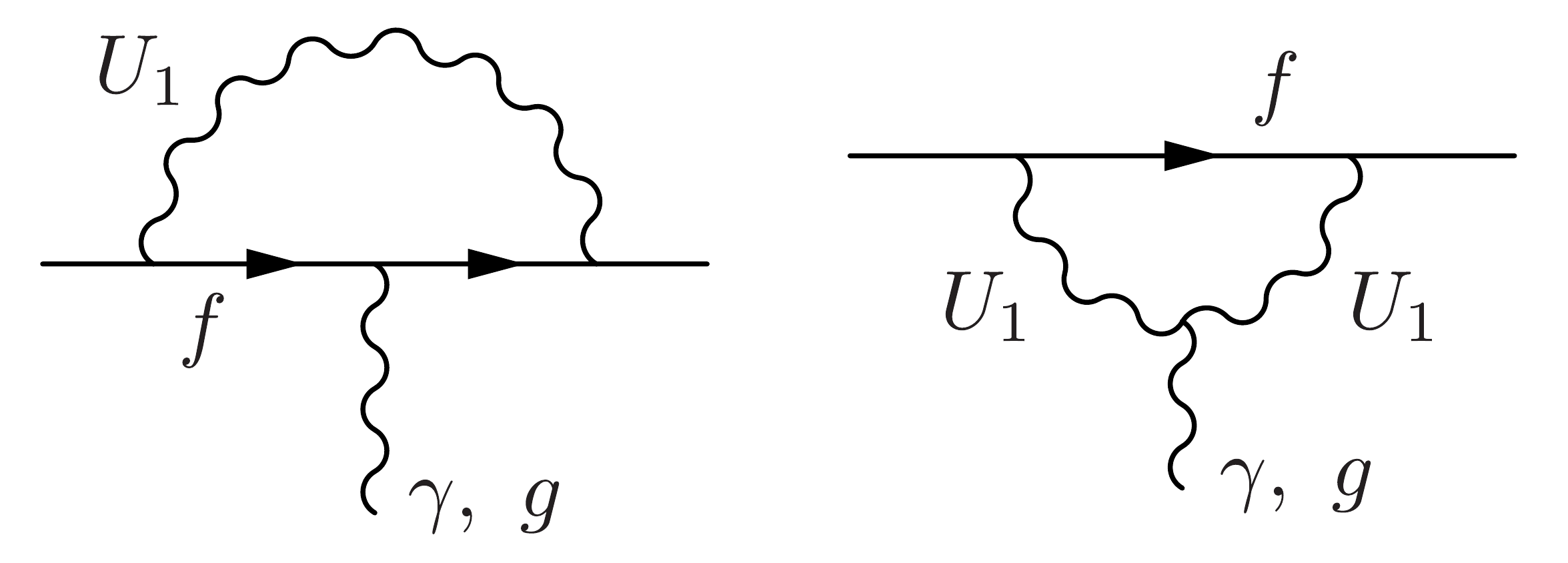}
\caption{Feynman diagrams contributing to the dipole moments of quarks and leptons from leptoquark exchange.}
\label{fig:diagrams}
\end{figure}

In this section, we calculate and present new and original formulae for shifts in the electric and magnetic dipole moments of leptons and quarks induced by the leptoquark.  We then estimate the size of the neutron electric dipole.  Finally, we review experimental limits on the dipole moments.

The leptoquark radiatively induces dipole moments starting at one loop order as shown in Fig.~\ref{fig:diagrams}.  After integrating out the leptoquark, effective interactions encoding the dipole moments are given by the effective Lagrangian
\begin{equation}
    \mathcal L_\text{eff} = \sum_f \left( a_f \frac{ e Q_f }{4 m_f} (\bar f \sigma^{\mu\nu} f) F_{\mu\nu} -\frac{i d_f}{2}  (\bar f \sigma^{\mu\nu} \gamma_5 f) F_{\mu\nu} \right)\,,
\end{equation}
 where $a_f$ is the anomalous magnetic dipole moment, and $d_f$ is the electric dipole moment of SM fermion $f$.  In the absence of right-handed neutrinos, the $U_1$ leptoquark does not generate dipole moments for neutrinos. 

Through its coupling with the gluons, the leptoquark induces chromomagnetic, $\hat a_q$, and chromoelectric, $\hat d_q$, dipole moments of quarks 
\begin{equation}
\mathcal L_\text{eff} = \sum_q \left(  \frac{\hat a_q}{4 m_q} (\bar q \sigma^{\mu\nu} T^a q) G^a_{\mu\nu} - \frac{i \hat d_q}{2} (\bar q \sigma^{\mu\nu} T^a \gamma_5 q) G^a_{\mu\nu} \right)\,.
\end{equation}

%%%%%%%%%%%%%%%%%%%%%%%%%%%%%%%%%%%%%%%%%%%%%%%%
\subsection{Leptoquark Contribution to Dipole Moments of SM Leptons and Quarks}

In the large $M_{U_1}$ limit, the leptoquark contribution to the anomalous magnetic moment of the muon is
\begin{multline} \label{eq:a_lepton}
a_\mu = \frac{N_C}{16\pi^2}\sum_i\Big[2\text{Re}( \lambda_{i2}^q  \lambda_{i2}^{d*})\frac{m_{d_i} m_\mu}{M_{U_1}^2}\Big(2 Q_d + Q_{U}\big((1-\kappa_Y)\ln(\frac{\Lambda_\text{UV}^2}{M_{U_1}^2})+\frac{1-5\kappa_Y}{2}\big)\Big) \\
+2Q_{U}{\tilde\kappa}_Y\text{Im}( \lambda_{i2}^q  \lambda_{i2}^{d*})\frac{m_{d_i} m_\mu}{M_{U_1}^2}\Big(\ln(\frac{\Lambda_\text{UV}^2}{M_{U_1}^2})+\frac{5}{2}\Big) \\
-(| \lambda_{i2}^q|^2 + | \lambda_{i2}^d|^2)\frac{m_\mu^2}{M_{U_1}^2}\Big(\frac{4}{3} Q_d + Q_{U}\big((1-\kappa_Y)\ln(\frac{\Lambda_\text{UV}^2}{M_{U_1}^2})-\frac{1+9\kappa_Y}{6}\big)\Big)\Big]\,,
\end{multline}
where $Q_d = -1/3$, $Q_U = +2/3$ is the leptoquark electric charge, and $N_C = 3$.  Note that if $\kappa_Y \neq 1$ or $\tilde \kappa_Y \neq 0$, relevant for scenarios in which the leptoquark is not a gauge boson, the dipole moment exhibits logarithmic dependence on the cut-off scale $\Lambda_{UV}$ not far above the leptoquark mass.  Our formula is in agreement with \cite{Queiroz:2014zfa,Kowalska:2018ulj} when specialized to the vector leptoquark model with $\kappa_Y=1$ and $\tilde\kappa_Y=0$.  Similarly, the muon electric dipole moment is
\begin{multline} \label{eq:d_lepton}
d_\mu = \frac{e N_C}{16\pi^2}\sum_i\Big[ \text{Im}( \lambda_{i2}^q  \lambda_{i2}^{d*})\frac{m_{d_i}}{M_{U_1}^2}\Big(2Q_d + Q_{U}\big((1-\kappa_Y)\ln(\frac{\Lambda_\text{UV}^2}{M_{U_1}^2})+\frac{1-5\kappa_Y}{2}\big)\Big) \\
 + Q_U {\tilde\kappa}_Y \text{Re}( \lambda_{i2}^q  \lambda_{i2}^{d*})\frac{m_{d_i}}{M_{U_1}^2}\big(\ln(\frac{\Lambda_\text{UV}^2}{M_{U_1}^2})+\frac{5}{2}\big) \\
 + Q_U {\tilde\kappa}_Y (| \lambda_{i2}^q|^2 + | \lambda_{i2}^d|^2)\frac{m_\mu}{M_{U_1}^2}\big(\frac{1}{2}\ln(\frac{\Lambda_\text{UV}^2}{M_{U_1}^2})+\frac{3}{4}\big)\Big]\,.
\end{multline}
CP violation is provided either by the imaginary part of the fermion coupling combination $ \lambda_{i2}^q  \lambda_{i2}^{d*}$, or by the CP violating hypercharge coupling $\tilde \kappa_Y$.  Dipole moments of other charged leptons are obtained by the appropriate replacement of the muon mass, $m_\mu$, and leptoquark couplings to muons, $\lambda_{i2}$. 

The bottom quark electric dipole moment induced by the leptoquark is
\begin{multline}\label{eq:strangequark-edm}
d_b = \frac{e}{16\pi^2}\sum_i\Big[\text{Im}( \lambda_{3i}^q  \lambda_{3i}^{d*})\frac{m_{\ell_i}}{M_{U_1}^2}\Big(2Q_\ell + Q_{U}\big((1-\kappa_Y)\ln(\frac{\Lambda_\text{UV}^2}{M_{U_1}^2})+\frac{1-5\kappa_Y}{2}\big)\Big) \\
 + Q_U {\tilde\kappa}_Y \text{Re}( \lambda_{3i}^q  \lambda_{3i}^{d*})\frac{m_{\ell_i}}{M_{U_1}^2}\big(\ln(\frac{\Lambda_\text{UV}^2}{M_{U_1}^2})+\frac{5}{2}\big) \\
 - Q_U {\tilde\kappa}_Y (| \lambda_{3i}^q|^2 + | \lambda_{3i}^d|^2)\frac{m_b}{M_{U_1}^2}\big(\frac{1}{2}\ln(\frac{\Lambda_\text{UV}^2}{M_{U_1}^2})+\frac{3}{4}\big)\Big]\,,
\end{multline}
and the chromoelectric dipole moment (cEDM) is
\begin{multline}\label{eq:strangequark-cedm}
\hat{d}_b = \frac{g_s}{16\pi^2}\sum_i \Big[ \text{Im}( \lambda_{3i}^q  \lambda_{3i}^{d*})\frac{m_{\ell_i}}{M_{U_1}^2}\Big((1-\kappa_Y)\ln(\frac{\Lambda_\text{UV}^2}{M_{U_1}^2})+\frac{1-5\kappa_Y}{2}\Big) \\
 + {\tilde\kappa}_Y \text{Re}( \lambda_{3i}^q  \lambda_{3i}^{d*})\frac{m_{\ell_i}}{M_{U_1}^2}\big(\ln(\frac{\Lambda_\text{UV}^2}{M_{U_1}^2})+\frac{5}{2}\big) \\
 - {\tilde\kappa}_Y (| \lambda_{3i}^q|^2 + | \lambda_{3i}^d|^2)\frac{m_b}{M_{U_1}^2}\big(\frac{1}{2}\ln(\frac{\Lambda_\text{UV}^2}{M_{U_1}^2})+\frac{3}{4}\big)\Big].
\end{multline}
The other down-type quark (chromo-)electric dipole moments can be obtained by appropriate replacements of flavor indices.

Analogously, up-type quark (chromo-)electric dipole moments are obtained from the bottom quark result by the replacement $ \lambda_{ij}^q \rightarrow V_{ik}  \lambda_{kj}^q$, $ \lambda_{ij}^d \rightarrow 0$, $m_\ell \rightarrow m_\nu = 0$, $m_b \rightarrow m_u$ yielding
\begin{equation}
d_u = - \frac{e}{16\pi^2} Q_U {\tilde\kappa}_Y \sum_i | (V\lambda^q)_{1i}|^2 \frac{m_u}{M_{U_1}^2}\Big(\frac{1}{2}\ln(\frac{\Lambda_\text{UV}^2}{M_{U_1}^2})+\frac{3}{4}\Big)\,,
\end{equation}

and
\begin{equation}
\hat {d}_u = - \frac{g_s}{16\pi^2} {\tilde\kappa}_Y \sum_i |(V\lambda^q)_{1i}|^2 \frac{m_u}{M_{U_1}^2}\Big(\frac{1}{2}\ln(\frac{\Lambda_\text{UV}^2}{M_{U_1}^2})+\frac{3}{4}\Big)\,.
\end{equation}
We do not consider anomalous (chromo-)magnetic moments of the quarks as they are experimentally not constrained.

\subsection{Connecting Quark Dipole Moments to the Neutron EDM}
In the following, we determine the neutron electric dipole moment due to quark-level dipole moments.  We neglect the running of quark dipole moments from the leptoquark scale to the hadronic scale, since the neglected logarithm of order $\alpha_s \ln(M_{U_1}^2/M_n^2) \approx 1.6$ leads to corrections which are small compared to the relevant hadronic uncertainties discussed below.  

The dominant contributions to the neutron EDM are from the short range QCD interactions involving quark EDMs, $d_i$, and cEDMs, $\hat{d}_i$,  given by
\begin{equation}\label{eq:shortRangeEDM}
d_n \sim -\frac{v}{\sqrt{2}}\Big[\beta^{uG}_n \hat{d}_u + \beta^{dG}_n \hat{d}_d + \beta^{sG}_n \hat{d}_s + \beta_n^{u\gamma} d_u + \beta_n^{d\gamma} d_d + \beta_n^{s\gamma} d_s\Big]\,,
\end{equation}
where the $\beta_i^{(k)}$ are the hadronic matrix elements.  Estimates from quark cEDM are given by $\beta^{uG}_n \approx 4^{+6}_{-3}\times10^{-4}\,e\text{ fm}$ and $\beta^{dG}_n \approx 8^{+10}_{-6}\times10^{-4}\,e\text{ fm}$ \cite{Engel:2013lsa}.  The most recent lattice evaluations of the matrix elements involving the electromagnetic EDMs are \cite{Bhattacharya:2015esa, Bhattacharya:2016zcn} $-\frac{v}{\sqrt{2}}\beta^{u\gamma}_n \approx -0.233(28)$, $-\frac{v}{\sqrt{2}}\beta^{d\gamma}_n \approx 0.776(66)$ and $-\frac{v}{\sqrt{2}}\beta^{s\gamma}_n \approx 0.008(9)$.

Contributions from heavy quark cEDM are estimated by integrating out the heavy quark, $Q=c,b$, to generate the three gluon Weinberg (gluon cEDM) operator,
\begin{equation}
\mathcal{L} = \frac{c_{\tilde{\text{G}}}}{m_Q^2} \frac{g_sf^{abc}}{3}\tilde{G}^a_{\mu\nu}G^b_{\nu\rho}G_{\rho}^{c\,\mu}\,,
\end{equation}
where the Wilson coefficient is given by \cite{Chang:1990jv,Boyd:1990bx,Dine:1990pf}
\begin{equation}
c_{\tilde{\text{G}}} = \frac{g_s^2}{32\pi^2} m_Q \hat{d}_Q\,.
\end{equation}
Contributions to $c_{\tilde{\text{G}}}$ from CP-violating leptoquark gluon interactions proportional to $\tilde{\kappa}_s$ are also present, but we do not consider them since they are unrelated to flavor anomalies.
In terms of $c_{\tilde{\text{G}}}$, the neutron EDM is given by \cite{Engel:2013lsa}
\begin{equation}
d_n = \frac{v^2}{m_Q^2} \beta^{\tilde{G}}_n c_{\tilde{\text{G}}}
\end{equation}
where $\beta^{\tilde{G}}_n\approx [2,\, 40]\times10^{-20} \text{ e cm}$ is the nucleon matrix element estimated using QCD sum rules and chiral perturbation theory \cite{Demir:2002gg}.

To compare the relative sizes of contributions from light and heavy quark to the neutron EDM, we take the strange and bottom quark contributions, and assume for simplicity that $\kappa_Y = 1$, $\tilde\kappa_Y = 0$.  We also assume $M_{U_1} \sim  2\text{ TeV}$ for the leptoquark scale.

Putting together Eqs. (\ref{eq:strangequark-edm}) and (\ref{eq:strangequark-cedm}) with Eq. (\ref{eq:shortRangeEDM}), we find that the strange quark EDM contribution to the neutron EDM is 

\begin{eqnarray}\nonumber
d_n^{\rm{strange}} &\approx& -\frac{5}{24\pi^2 M_{U_1}^2 \text{ cm}}\left[m_\tau \text{Im}( \lambda_{23}^q  \lambda_{23}^{d*}) +m_\mu \text{Im}( \lambda_{22}^q  \lambda_{22}^{d*})\right]\times 0.008  \text{ $e$ cm}\\
\label{eq:s-neturon-edm}&\sim& -\left(\text{Im}( \lambda_{23}^q  \lambda_{23}^{d*})+0.06~\text{Im}( \lambda_{22}^q  \lambda_{22}^{d*})\right)\times 1.5\times 10^{-24}\text{ $e$ cm}\,.
\end{eqnarray}
The bottom quark cEDM contribution to the neutron EDM is instead given by

\begin{eqnarray}\nonumber
d_n^{\rm{bottom}} &\approx& -\frac{g_s^3  v^2}{(16\pi^2)^2 m_b M_{U_1}^2}\left[m_\tau \text{Im}( \lambda_{33}^q  \lambda_{33}^{d*})+m_\mu \text{Im}( \lambda_{32}^q  \lambda_{32}^{d*})\right] \times[2, 40]\times 10^{-20}\\
\label{eq:b-neturon-edm}&\sim& -\left(\text{Im}( \lambda_{33}^q  \lambda_{33}^{d*})+0.06~\text{Im}( \lambda_{32}^q  \lambda_{32}^{d*})\right)\times [2,40]\times5\times 10^{-27}\text{ $e$ cm}\,.
\end{eqnarray}
For generic $\mathcal{O}(1)$ sized leptoquark couplings $\lambda^q_{ik}$ and $\lambda^d_{ik}$ the strange quark contribution (\ref{eq:s-neturon-edm}) to the neutron EDM is much larger than the bottom quark contribution (\ref{eq:b-neturon-edm}).  However, in the region of parameter space we are exploring, the bottom quark contribution is typically bigger than the strange quark contribution. 

\subsection{Experimental Status and Prospects}
We review here the current experimental status of dipole moments of Standard Model fermions. 
The anomalous magnetic moments of the electron, $a_e$, and the muon, $a_\mu$, are measured extremely precisely~\cite{Hanneke:2008tm,Bennett:2006fi}, and are predicted to similarly high precision within the SM, with new physics contributions constrained to lie within the range~\cite{Keshavarzi:2019abf,Davoudiasl:2018fbb} (see also~\cite{Davier:2019can,Crivellin:2018qmi,Jegerlehner:2017lbd})
\begin{equation}
 \Delta a_\mu = (28.0 \pm 6.3_{\rm{exp}}\pm 3.8_{\rm{th}}) \times 10^{-10} \,, \qquad \Delta a_e  = (-8.9 \pm 3.6_{\rm{exp}}\pm 2.3_{\rm{th}} ) \times 10^{-13} \,,
\end{equation}
 In addition to the long standing discrepancy in the muon magnetic moment with a significance of more than $3\sigma$, a discrepancy in the electron magnetic moment arose after a recent precision measurement of the fine structure constant~\cite{Parker:2018vye} with a significance of $\sim 2.4\sigma$. Combining the expected sensitivity from the running $g-2$ experiment at Fermilab~\cite{Grange:2015fou} with expected progress on the SM prediction (see~\cite{Blum:2016lnc,Blum:2018mom,Davies:2019efs,Shintani:2019wai,Gerardin:2019rua,Blum:2019ugy} for recent lattice efforts and \cite{Colangelo:2015ama,Colangelo:2017qdm,Hoferichter:2018dmo,Colangelo:2018mtw,Hoferichter:2019gzf} for recent efforts using the framework of dispersion relations) the uncertainty on $\Delta a_\mu$ will be reduced by a factor of a few in the coming years. Similarly, for $\Delta a_e$ we expect an order of magnitude improvement in the sensitivity~\cite{Gabrielse:2019cgf}.

The anomalous magnetic moment of the tau, $a_\tau$, is currently only very weakly constrained. The strongest constraint comes from LEP and reads at 95\% C.L.~\cite{Abdallah:2003xd}
\begin{equation}
 -0.055 < a_\tau < 0.013 \,.
\end{equation}
Improvements in sensitivity by an order of magnitude or more might be achieved at Belle II or future electron positron colliders (see~\cite{Pich:2013lsa} for a review).

Strong experimental constraints exist for the EDM of the electron. The strongest bound is inferred from the bound on the EDM of ThO obtained by the ACME collaboration which gives at 90\% C.L.~\cite{Andreev:2018ayy}
\begin{equation}
 |d_e| < 1.1 \times 10^{-29} \, e\,\text{cm}\,.
\end{equation}
Significant improvements by an order of magnitude or more can be expected from ACME in the future~\cite{Andreev:2018ayy}.

Only weak constraints exist for the EDMs of the muon and the tau, $d_\mu$ and $d_\tau$. Analyses by the Muon g-2 collaboration~\cite{Bennett:2008dy} and the Belle collaboration~\cite{Inami:2002ah} give the following bounds at 95\% C.L.
\begin{equation}
 |d_\mu| < 1.9 \times 10^{-19} \, e\,\text{cm}\,, \qquad -2.2 \times 10^{-17} \, e\,\text{cm} < d_\tau < 4.5 \times 10^{-17}  \,e\,\text{cm}\,.
\end{equation}
The proposed muon EDM experiment at PSI aims at improving the sensitivity to the muon EDM by 4 orders of magnitude, $d_\mu \lesssim 5 \times 10^{-23}  e$\,cm~\cite{Adelmann:2010zz}. Improving the sensitivity to the tau EDM by roughly two orders of magnitude ($d_\tau<2\times 10^{-19}~e$\,cm) might be possible at Belle II or at future $e^+e^-$ colliders~\cite{Kou:2018nap}.

%
%
%The proposed Muon g-2/EDM experiment at J-PARC aims at improving the sensitivity to the muon EDM by 2 orders of magnitude, $d_\mu \lesssim 1.5 \times 10^{-21}  e$\,cm~\cite{Abe:2019thb}.
%Improving the sensitivity to the tau EDM by roughly two orders of magnitude ($d_\tau<2\times 10^{-19}~e$ cm) might be possible at Belle II or at future $e^+e^-$ colliders~\cite{Kou:2018nap}.

Turning to quarks, we note that the magnetic and chromo-magnetic dipole moments of quarks, $a_q$ and $\hat a_q$, are very weakly constrained and we therefore do not consider them in this work.
As discussed in the previous section, the EDMs and cEDMs of quarks, $d_q$ and $\hat d_q$, lead to EDMs of hadronic systems like the neutron and are therefore strongly constrained. In the following we will focus on the neutron EDM which is bounded at 95\% C.L. by~\cite{Afach:2015sja}
\begin{equation}
 |d_n| < 3.6 \times 10^{-26} \, e\,\text{cm}\,.
\end{equation}
Experimental sensitivities should improve by two orders of magnitude to a few $10^{-28}e\,\text{cm}$ in the next decade \cite{Strategy:2019vxc}.

We collect the SM predictions, the current experimental results, and expected future experimental sensitivities to the dipole moments in Table~\ref{tab:expstatus}.
%%%%%%%%%%%%%%%%%%%%%%%%%%%%%%%%%
\begin{table}[tb]
\begin{ruledtabular}
\begin{tabular}{cccc}
observable & SM theory & current exp. & projected sens. \\
\hline
$a_e-a_e^{\text{SM}}$ & $\pm 2.3\times{10^{-13}}$ \cite{Keshavarzi:2019abf,Parker:2018vye}& $(-8.9\pm3.6)\times 10^{-13}$ \cite{Hanneke:2008tm} & $\sim10^{-14}$ \cite{Gabrielse:2019cgf}
\\
$a_\mu - a_\mu^{\text{SM}}$ & $\pm3.8\times10^{-10}$  \cite{Keshavarzi:2019abf} & $(28.0\pm 6.3)\times10^{-10}$ \cite{Bennett:2006fi}&$ 1.6\times10^{-10}$ \cite{Grange:2015fou}\\
$a_\tau - a_\tau^\text{SM}$ &$\pm3.9 \times 10^{-8}$  \cite{Keshavarzi:2019abf} & $(-2.1\pm1.7)\times10^{-2}$ \cite{Abdallah:2003xd}& \\
\hline
$d_e$ & 
 $< 10^{-44}~e$ cm \cite{Pospelov:2013sca,Smith:2017dtz} & $< 1.1 \times 10^{-29} \,e\,\text{cm}$ \cite{Andreev:2018ayy} & $ \sim 10^{-30} \,e\,\text{cm}$ \cite{Andreev:2018ayy}\\
$d_\mu$ &$< 10^{-42}~e$ cm \cite{Smith:2017dtz} & $ < 1.9 \times 10^{-19} \,e\,\text{cm}$ \cite{Bennett:2008dy} & $ \sim 10^{-23} \,e\,\text{cm}$ \cite{Adelmann:2010zz} \\
$d_\tau$ &$< 10^{-41}~e$ cm \cite{Smith:2017dtz} & $ (1.15 \pm 1.70) \times 10^{-17} \,e\,\text{cm}$ \cite{Inami:2002ah} & $ \sim 10^{-19} \,e\,\text{cm}$ \cite{Kou:2018nap} \\
\hline
$d_n$ & $\sim 10^{-32}~e$ cm \cite{Dar:2000tn}& $< 3.6 \times 10^{-26} \, e\,\text{cm}$~\cite{Afach:2015sja} & few$\times 10^{-28}e\,\text{cm}$ \cite{Strategy:2019vxc}\\
\end{tabular}
\end{ruledtabular}
\caption{Summary of Standard Model theory errors/bounds (first column), current experimental measurements/limits (second column) and projected precision of next-generation experiments (third column) of magnetic moment anomalies and electric dipole moments of the charged leptons and the neutron.  For clarity, for the anomalous magnetic moments, the Standard Model central values have been subtracted. We are not aware of any experimental analysis for the projected sensitivity of the tau magnetic moment.}
\label{tab:expstatus}
\end{table}
%%%%%%%%%%%%%%%%%%%%%%%%%%%%%%%%%

%%%%%%%%%%%%%%%%%%%%%%%%%%%%%%%%%%%%%%%%%%%%%
\section{Flavor Anomalies and Electric Dipole Moments}\label{Sec:Pheno}
%%%%%%%%%%%%%%%%%%%%%%%%%%%%%%%%%%%%%%%%%%%%%
In this section, we study the impact of leptoquarks on (c)EDMs and $B$-physics measurements at the benchmark points presented in Sec. \ref{sec:LQBdecays}.

%%%%%%%%%%%%%%%%%%%%%%%%%%%%%%%%%%%%%%%%%%%%%
\subsection{Probing the Parameter Space Using Tau Measurements}\label{sec:flavor-edm-pheno}
%%%%%%%%%%%%%%%%%%%%%%%%%%%%%%%%%%%%%%%%%%%%%

Given the BM1 and BM2 benchmarks for the leptoquark couplings to left-handed taus, $\lambda^q_{33} \simeq 0.7$, $\lambda^q_{23} \simeq 0.6$, we begin by turning on the coupling to right-handed taus $\lambda^d_{33}$ while setting the right-handed couplings to muons and electrons ($\lambda^d_{32}$  and $\lambda^d_{31}$, respectively) to zero. The coupling $\lambda^d_{33}$ will induce the dipole moments of the tau as in Eqs.~\eqref{eq:a_lepton} and~\eqref{eq:d_lepton}, as well as transition dipole moments leading to the lepton flavor violating decay modes $\tau \to \mu \gamma$ and $\tau \to e \gamma$. In the limit $m_e, m_\mu \ll m_\tau \ll m_b$, the partial width for the $U_1$ contribution to $\tau \to \mu \gamma$ is given by
\begin{multline} \label{eq:taumugamma}
\Gamma_{\tau\rightarrow\mu\gamma} = \frac{\alpha m_\tau^3 N_C^2}{256\pi^4 M_{U_1}^4} m_b^2 |\lambda_{32}^q \lambda_{33}^{d*}|^2 \Bigg[ \Big(2 Q_b - Q_{U}\big((1-\kappa_Y)\ln(\frac{\Lambda_\text{UV}^2}{M_{U_1}^2})+\frac{1-5\kappa_Y}{2}\big)\Big)^2 \\ 
+ Q_{U}^2 \tilde{\kappa}_Y^2 \big(\ln(\frac{\Lambda_\text{UV}^2}{M_{U_1}^2})+\frac{5}{2}\big)^2\Bigg].
\end{multline}
This expression is in agreement with \cite{Cornella:2019hct}, when specialized to the vector leptoquark model with $\kappa_Y=1$ and $\tilde\kappa_Y = 0$. 
The expression for the decay mode $\tau \to e \gamma$ is obtained by an appropriate replacement of the lepton flavor index. 
The experimental upper limits on the branching ratios of the $\tau \to \mu \gamma$ and $\tau \to e \gamma$ decays are $5.0 \times 10^{-8}$ and $5.4 \times 10^{-8}$, respectively~\cite{Amhis:2019ckw}.

In addition to inducing lepton flavor violating tau decays, the $\lambda^d_{33}$ coupling will modify the new physics contributions to charged current decays based on the $b \to c \tau \nu$ and $b \to u \tau \nu$ transitions and neutral current decays based on $b \to s \tau \tau$. The decay modes that are particularly sensitive to right-handed currents are the helicity suppressed two body decays $B_c \to \tau \nu$~\cite{Li:2016vvp,Alonso:2016oyd}, $B^\pm \to \tau \nu$, and $B_s \to \tau^+ \tau^-$. We find
\begin{eqnarray} \label{eq:Bctaunu}
\frac{\text{BR}(B_c \to \tau \nu)}{\text{BR}(B_c \to \tau \nu)_\text{SM}} &=& \left| 1 - \frac{\sum_{j}V_{cj}\lambda^q_{j3}}{V_{cb}} \frac{v^2}{M_{U_1}^2} \left(\frac{\lambda^{q\ast}_{33}}{2}+ \frac{\lambda^{d\ast}_{33} m_{B_c}^2}{m_\tau(m_b+m_c)} \right) \right|^2 \,, \\ \label{eq:Btaunu}
\frac{\text{BR}(B^\pm \to \tau \nu)}{\text{BR}(B^\pm \to \tau \nu)_\text{SM}} &=& \left| 1 - \frac{\sum_j V_{uj} \lambda^q_{j3} }{V_{ub}} \frac{v^2}{M_{U_1}^2} \left(\frac{\lambda^{q*}_{33}}{2}+ \frac{\lambda^{d*}_{33} m_{B^\pm}^2}{m_\tau m_b} \right) \right|^2 \,,
\end{eqnarray}
Using the expression for the branching ratio in terms of the Wilson coefficients from~\cite{Altmannshofer:2017wqy}, we find
\begin{eqnarray} \label{eq:Bstautau}
\frac{\text{BR}(B_s \to \tau^+ \tau^-)}{\text{BR}(B_s \to \tau^+ \tau^-)_\text{SM}} &=& \left| 1 + \frac{4\pi^2}{e^2 C_{10}^\text{SM}} \frac{v^2}{M_{U_1}^2} \left(\frac{\lambda_{33}^{q*} \lambda_{23}^{q} + \lambda_{33}^{d*} \lambda_{23}^d}{V_{ts}^* V_{tb}} - \frac{m_{B_s}^2}{m_\tau m_b} \frac{\lambda_{33}^{q*} \lambda_{23}^d + \lambda_{33}^{d*} \lambda_{23}^q}{V_{ts}^* V_{tb}}\right) \right|^2  \nonumber \\
&&  + \frac{16\pi^4}{e^4 (C_{10}^\text{SM})^2} \frac{v^4}{M_{U_1}^4} \frac{m_{B_s}^4}{m_\tau^2 m_b^2} \left|\frac{\lambda_{33}^{q*} \lambda_{23}^d - \lambda_{33}^{d*} \lambda_{23}^q}{V_{ts}^* V_{tb}} \right|^2 \left( 1 - \frac{4 m_\tau^2}{m_{B_s}^2} \right) \,,
\end{eqnarray}
where we neglected the finite life time difference in the $B_s$ system. We use a normalization such that the SM value for the Wilson coefficient is $C_{10}^\text{SM} \simeq -4.1$~\cite{Altmannshofer:2008dz}. Renormalization group running from the leptoquark scale down to the $b$-scale can be incorporated by evaluating the quark masses in Eqs.~\eqref{eq:Bctaunu}-\eqref{eq:Bstautau} at the scale $\mu \simeq 2$\,TeV. Note that the terms containing both left-handed and right-handed couplings enjoy a mild chiral enhancement by factors $m_{B_c}^2/(m_\tau (m_b+m_c))$, $m_{B^\pm}^2/(m_\tau m_b)$, and $m_{B_s}^2/(m_\tau m_b)$, respectively. 

The measured BR$(B^\pm \to \tau \nu) = (1.09 \pm 0.24)\times 10^{-4}$~\cite{Tanabashi:2018oca} agrees well with the SM prediction BR$(B^\pm \to \tau \nu)_\text{SM} = (0.838^{+0.039} _{-0.029})\times 10^{-4}$~\cite{Charles:2004jd}, yielding
\begin{equation}
\frac{\text{BR}(B^\pm \to \tau \nu)}{\text{BR}(B^\pm \to \tau \nu)_\text{SM}} = 1.30 \pm 0.29 \,.
\end{equation}

So far no direct measurement of the $B_c \to \tau \nu$ branching ratio has been performed. We impose the bound BR$(B_c \to \tau \nu) < 30\%$~\cite{Alonso:2016oyd}.
The SM branching ratio is
\begin{equation}
\text{BR}(B_c \to \tau \nu)_\text{SM} = \tau_{B_c} m_{B_c}  \frac{f_{B_c}^2 G_F^2}{8\pi} |V_{cb}|^2 m_\tau^2 \left(1 - \frac{m_\tau^2}{m_{B_c}^2} \right)^2= (2.21 \pm 0.09) \times 10^{-2}\,,
\end{equation}
with the lifetime of the $B_c$ meson $\tau_{B_c}=(0.507 \pm 0.009)\times 10^{-12}$\,s~\cite{Tanabashi:2018oca}, the $B_c$ decay constant $f_{B_c} = (0.427\pm0.006)$\,GeV~\cite{McNeile:2012qf} and we used $|V_{cb}| = (41.6 \pm 0.56 )\times 10^{-3}$~\cite{Charles:2004jd}.

Similarly, the $B_s \to \tau^+ \tau^-$ decay has not been observed so far. The first direct limit on the branching ratio was placed by LHCb~\cite{Aaij:2017xqt} and is $\text{BR}(B_s \to \tau^+ \tau^-) < 6.8 \times 10^{-3}$, while the SM branching ratio is  $\text{BR}(B_s \to \tau^+ \tau^-)_\text{SM} = (7.73\pm 0.49)\times 10^{-7}$ \cite{Bobeth:2013uxa}. 

\begin{figure}
\centering
\includegraphics[width=0.5\textwidth]{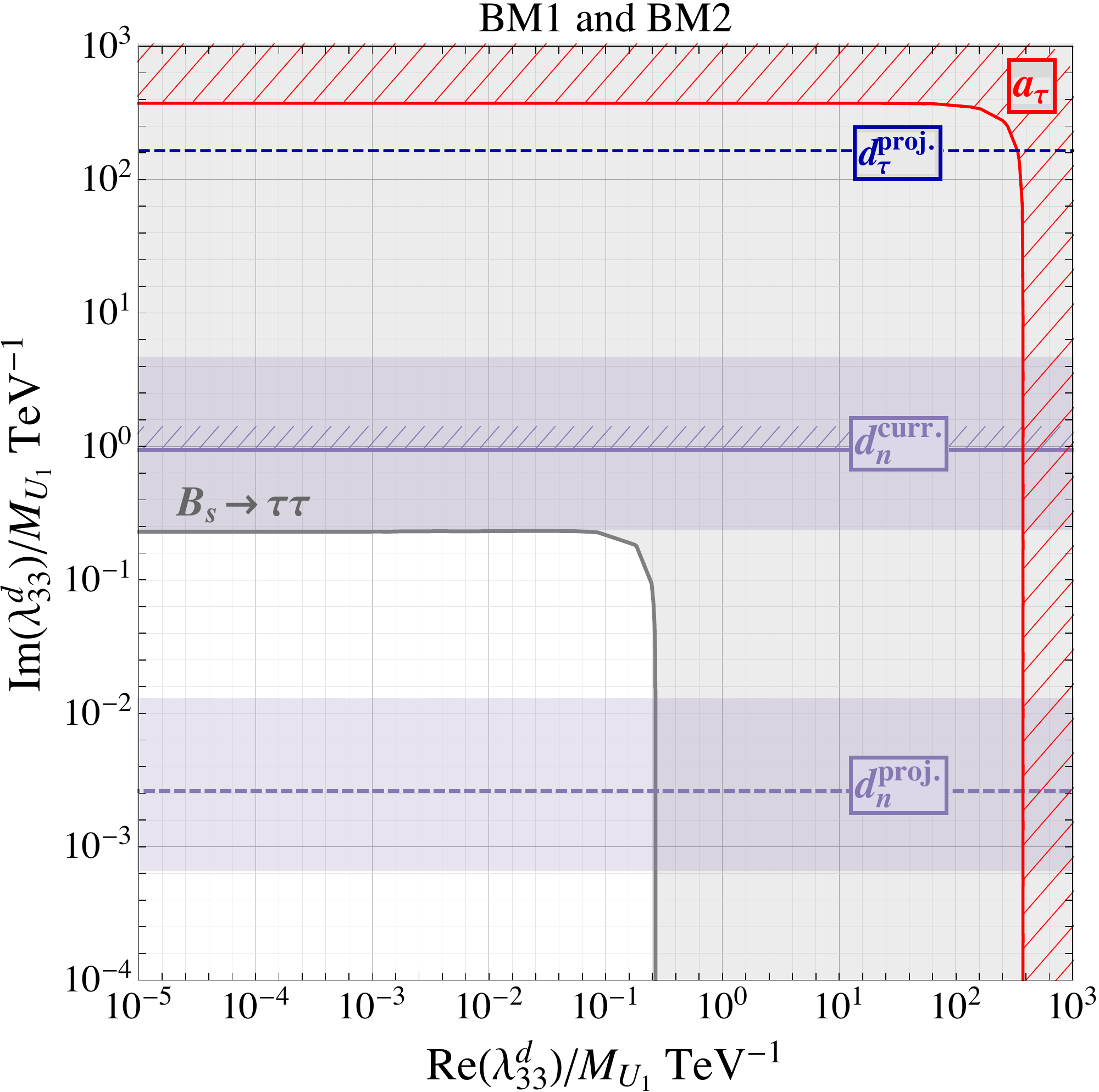}
\caption{Constraints on the $U_1$ leptoquark parameter space in the plane of the complex coupling $\lambda^d_{33}$ divided by the leptoquark mass, $M_{U_1}$, and all other parameters fixed as in BM1 (\ref{eq:BM1}) or BM2 (\ref{eq:BM2}). The gray region represents parameter space that is excluded by  $B_s \to \tau^+ \tau^-$. The red hatched region is excluded by the bound on the tau lepton anomalous magnetic moment. The dashed blue line is the projected sensitivity of future experiments to the tau EDM. The region above the solid purple line is excluded by bounds on the neutron EDM, and the dashed purple line is the projected sensitivity of future neutron EDM experiments. The surrounding purple bands reflect the theoretical uncertainty in the nucleon matrix element $\beta^{\tilde{G}}_n$. Note that the observables shown in the figure are independent of $\lambda^q_{32},\lambda^q_{22}$ and $\lambda^q_{31},\lambda^q_{21}$, and the change from benchmark BM1 to BM2 has no effect on the exclusion curves.}
\label{fig:tau}
\end{figure}

In Fig.~\ref{fig:tau}, we show current and projected constraints on the $U_1$ leptoquark in the plane of the complex $\lambda^d_{33}$ coupling divided by the leptoquark mass for BM1 and BM2 benchmark points. The figure represents both BM1 and BM2, since the shown constraints are independent of the muon couplings $\lambda^q_{32},\lambda^q_{22} $ and electron couplings $\lambda^q_{31},\lambda^q_{21}$ and changing from BM1 to BM2 does not affect our results.
The most stringent constraint comes from $B_s \to \tau^+ \tau^-$ and is shown in gray in the figure. Constraints from $B^\pm \to \tau \nu$, $B_c \to \tau \nu$, and lepton flavor violating tau decays ($\tau \to \mu \gamma$ for benchmark BM1 and $\tau \to e \gamma$ for BM2) are slightly weaker and exclude values of $\lambda^d_{33}$ that are a factor of a few larger than those excluded by $B_s \to \tau^+ \tau^-$. (In Fig.~\ref{fig:tau} we show only the strongest constraint coming from $B_s \to \tau^+ \tau^-$.) Once the bounds are imposed, the allowed values of the right-handed coupling $\lambda^d_{33}$ are sufficiently small such that they do not affect $R_{D^{(*)}},~R_{K^{(*)}}$ in a significant way. Therefore, in all the allowed region in Fig.~\ref{fig:tau}, the anomalies are satisfied.

From the figure, we observe that the current experimental bounds on $d_\tau$ and $a_\tau$ do not constraint the parameter space in a relevant way. The constraint from $a_\tau$ is depicted by the red hatched region in Fig.~\ref{fig:tau}, while the experimental bound on $d_\tau$ constrains values of $\Im(\lambda^d_{33})/M_{U_1}$ that are $\mathcal{O}(10^5)$~TeV$^{-1}$, and, therefore, beyond the range of the plot. Projected sensitivities of next-generation experiments to the tau EDM \cite{Kou:2018nap} (shown by the dashed blue line) are still far from being able to probe the viable new physics parameter space. 

In addition to the tau electric and anomalous magnetic dipole moments, the $U_1$ leptoquark coupling, $\lambda^d_{33}$, will contribute to the neutron EDM, $d_n$. The constraint from the current bound on the neutron EDM is shown by the solid purple line in Fig.~\ref{fig:tau}, where the region above this line is excluded due to the leptoquark generating a contribution to the neutron EDM that is too large. The surrounding purple bands reflect the theoretical uncertainty in the nucleon matrix element $\beta^{\tilde{G}}_n$. We observe that the currend bound on the neutron EDM leads to a constraint that is weaker than $B_s \to \tau^+ \tau^-$ and is not yet probing the allowed parameter space. On the other hand, the projected sensitivity of future neutron EDM experiments \cite{Strategy:2019vxc} (shown by the dashed purple line) will begin probing the new physics parameter space and can lead to stronger constraints on the amount of CP violation present in the right-handed couplings of $U_1$ to tau leptons.

%%%%%%%%%%%%%%%%%%%%%%%%%%%%%%%%%%%%%%%%%%%%%
\subsection{Probing the Parameter Space Using Muon Measurements} \label{sec:muon}
%%%%%%%%%%%%%%%%%%%%%%%%%%%%%%%%%%%%%%%%%%%%%

Next we focus on the BM1 and BM3 benchmarks, and investigate the impact of the leptoquark couplings to right-handed muons, $\lambda^d_{32}$, while setting the right-handed tau and electron couplings ($\lambda^d_{33}$ and $\lambda^d_{31}$, respectively) to zero.
The coupling $\lambda^d_{32}$ will lead to a shift in the anomalous magnetic moment of the muon, $\Delta a_\mu$, in the muon EDM, $d_\mu$, and in the EDM of the bottom quark given in Eqs.~\eqref{eq:a_lepton},~\eqref{eq:d_lepton}, and~\eqref{eq:strangequark-edm}, as well as the lepton flavor violating decay mode $\tau\to\mu\gamma$ given in  Eq.~\eqref{eq:taumugamma} with $|\lambda_{32}^q \lambda_{33}^{d*}|^2\rightarrow|\lambda_{32}^d \lambda_{33}^{q*}|^2$.  In the presence of the coupling  $\lambda^d_{32}$, the muon dipole moment enjoys a sizable chiral enhancement by $m_b/m_\mu$. 

In addition, the coupling $\lambda^d_{32}$ can also give sizable non-standard effects in the $B_s \to \mu^+\mu^-$ decay. The corresponding expression is analogous to the one for the $B_s \to \tau^+\tau^-$ decay given in Eq.~(\ref{eq:Bstautau})
\begin{eqnarray} \label{eq:Bsmumu}
\frac{\text{BR}(B_s \to \mu^+ \mu^-)}{\text{BR}(B_s \to \mu^+ \mu^-)_\text{SM}} &=& \left| 1 + \frac{4\pi^2}{e^2 C_{10}^\text{SM}} \frac{v^2}{M_{U_1}^2} \left(\frac{\lambda_{32}^{q*} \lambda_{22}^{q} + \lambda_{32}^{d*} \lambda_{22}^d}{V_{ts}^* V_{tb}} - \frac{m_{B_s}^2}{m_\mu m_b} \frac{\lambda_{32}^{q*} \lambda_{22}^d + \lambda_{32}^{d*} \lambda_{22}^q}{V_{ts}^* V_{tb}}\right) \right|^2  \nonumber \\
&&  + \frac{16\pi^4}{e^4 (C_{10}^\text{SM})^2} \frac{v^4}{M_{U_1}^4} \frac{m_{B_s}^4}{m_\mu^2 m_b^2} \left|\frac{\lambda_{32}^{q*} \lambda_{22}^d - \lambda_{32}^{d*} \lambda_{22}^q}{V_{ts}^* V_{tb}} \right|^2 \,.
\end{eqnarray}
The terms that contain both left-handed and right-handed couplings are chirally enchanced by a factor $m_{B_s}^2/(m_\mu m_b)$. 

The branching ratio BR$(B_s \to \mu^+ \mu^-)$ has been measured at LHCb, CMS and ATLAS~\cite{Chatrchyan:2013bka,CMS:2014xfa,Aaij:2017vad,Aaboud:2018mst}. We use the average of these results from~\cite{Aebischer:2019mlg}, that, combined with the SM prediction~\cite{Bobeth:2013uxa,Beneke:2017vpq}, reads
\begin{equation}
\frac{\text{BR}(B_s \to \mu^+ \mu^-)}{\text{BR}(B_s \to \mu^+ \mu^-)_\text{SM}} = 0.73 ^{+0.13} _{-0.10}\,,
\end{equation}
which is in slight tension ($\sim 2\sigma$) with the SM prediction. Interestingly enough, in the region of parameter space where the couplings to left-handed muons $\lambda^q_{22}$, $\lambda^q_{32}$ provide an explanation of $R_{K^{(*)}}$, the tension in $B_s \to \mu^+ \mu^-$ is largely lifted.

\begin{figure}
\centering
\includegraphics[width=0.45\textwidth]{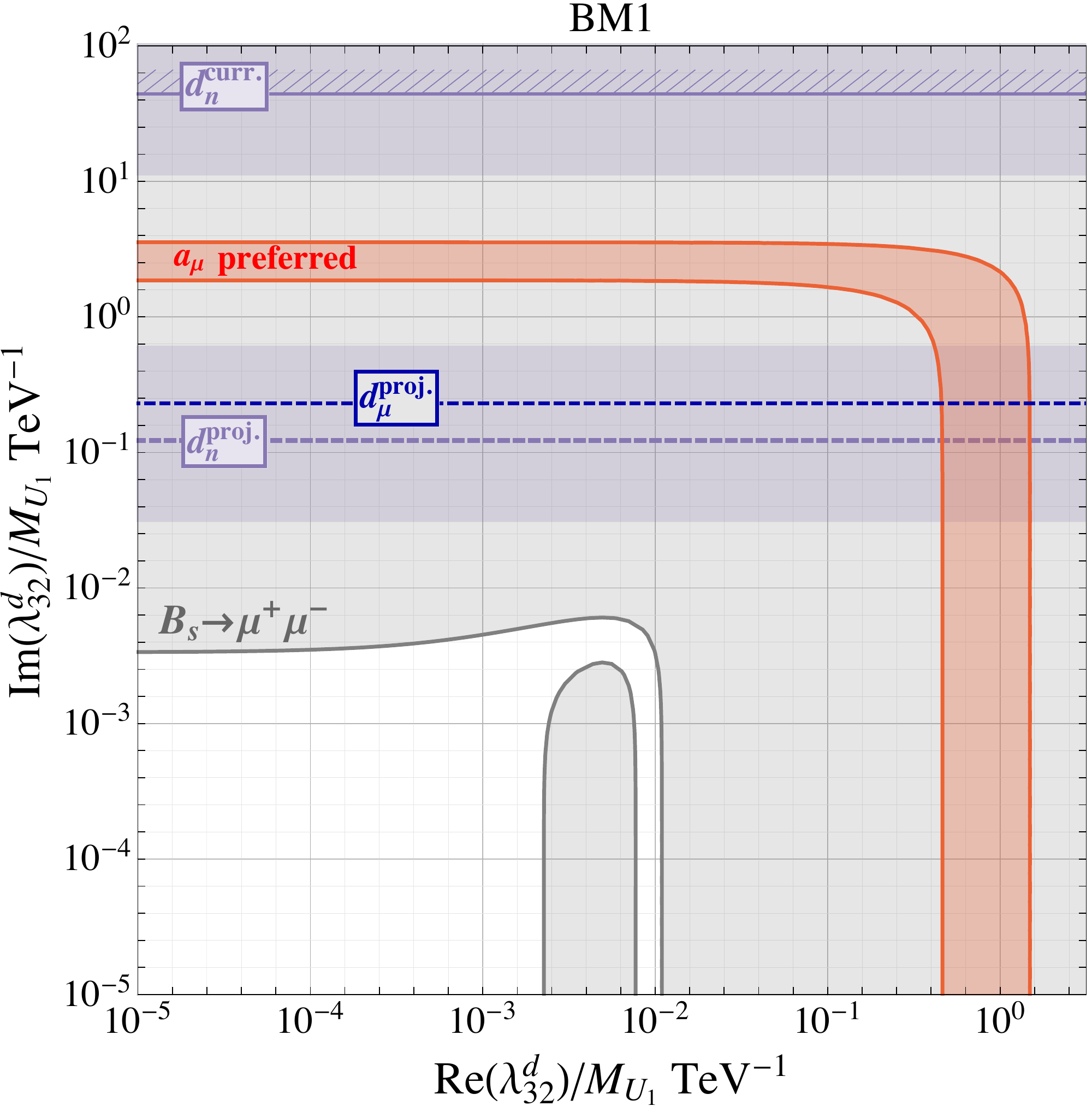} ~~~~
\includegraphics[width=0.45\textwidth]{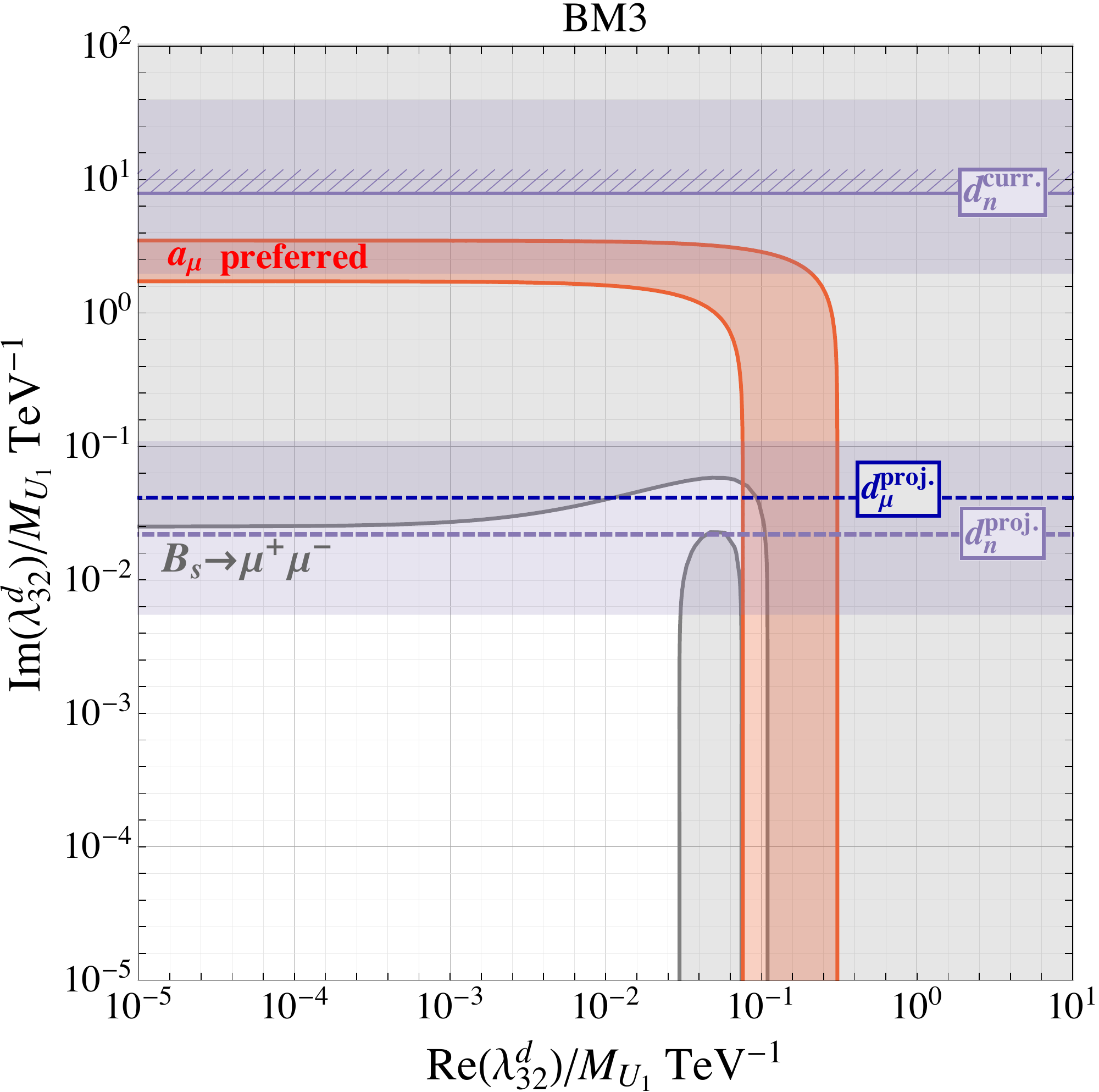}
\caption{Constraints on the $U_1$ leptoquark parameter space in the plane of the complex $\lambda^d_{32}$ coupling divided by the leptoquark mass for the benchmark points BM1 (left panel) and BM3 (right panel). The gray region is excluded by $B_s \to \mu^+ \mu^-$ at the 95\% C.L.. The dashed blue line is the projected sensitivity of future experiments to the muon EDM. The red shaded region corresponds to the parameter space the can address the anomaly in the anomalous magnetic moment of the muon. The solid (dashed) purple lines represent the current constraint (projected sensitivity) from the neutron EDM, with the purple bands reflecting the uncertainty in the nucleon matrix element $\beta^{\tilde{G}}_n$.}
\label{fig:muon}
\end{figure}

In Fig.~\ref{fig:muon} we show the current and projected constraints on the $U_1$ leptoquark for BM1 (left) and BM3 (right) in the plane of the complex coupling $\lambda^d_{32}$ divided by the leptoquark mass. For both benchmarks, the most stringent constraint arises from $B_s \to \mu^+\mu^-$. The region that is excluded at the $95\%$ C.L. is shaded in gray. Once the constraints from $B_s \to \mu^+ \mu^-$ are imposed, the allowed values of $\lambda^d_{32}$ are sufficiently small that they do not affect $R_{K^{(\ast)}}$ in a significant way. The region that is shaded in red is the region of parameter space that is able to address the anomaly in the anomalous magnetic moment of the muon, while the blue dashed lines are the projected sensitivities of future experiments to the muon EDM. Similar to Fig.~\ref{fig:tau}, the solid (dashed) purple line is the current constraint (projected sensitivity) of the neutron EDM. The current bound on the muon EDM, $d_\mu$, is very weak and constrains values of $\Im (\lambda^d_{32})/M_{U_1}$ outside from the range of the plot ($\Im (\lambda^d_{32})/M_{U_1}\sim \mathcal{O}(10^3)$~TeV$^{-1}$ for BM1 and $\Im(\lambda^d_{32})/M_{U_1}\sim \mathcal{O}(10^2)$~TeV$^{-1}$ for BM3). 

In the left plot of Fig.~\ref{fig:muon} we observe that, once the constraints from $B_s \to \mu^+ \mu^-$ is imposed, the BM1 benchmark cannot address the $a_\mu$ anomaly. We conclude that the $U_1$ leptoquark can not explain the $B$ anomalies and the $(g-2)_\mu$ anomaly simultaneously with the parameters fixed to those of BM1. This is mainly due to limits on lepton flavor violating decays $\tau \to \phi \mu$ and $B \to K \tau \mu$ that impose stringent constraints on the size of the left-handed muonic couplings $\lambda^q_{32}$ and $\lambda^q_{22}$ (see discussion in Sec.~\ref{sec:LQBdecays}). 

In order to avoid these constraints, we can instead set the $U_1$ couplings to left-handed tau leptons, $\lambda^q_{33}$ and $\lambda^q_{23}$, to zero as in BM3 in (\ref{eq:BM3}). The decay rates $\tau \to \phi \mu$, $B \to K \tau \mu$, and $\tau \to \mu \gamma$ mediated by $U_1$ then go to zero, allowing the muonic couplings $\lambda^q_{32}$ and $\lambda^q_{22}$ to have larger values.
However, by switching off $\lambda^q_{33}$ and $\lambda^q_{23}$ we forgo an explanation of $R_{D^{(\ast)}}$. 

In the right plot of Fig.~\ref{fig:muon} we show that, for BM3, the region of parameter space that can address the $a_\mu$ anomaly (the red shaded region) overlaps with the region of parameter space that is allowed by $B_s \to \mu^+ \mu^-$, and the $U_1$ leptoquark can therefore address both the $(g-2)_\mu$ anomaly and (at least partially, cf. discussion in Sec.~\ref{sec:LQBdecays}) the $R_{K^{(\ast)}}$ anomalies.
Finally, we notice that, for this benchmark, projected sensitivities to the neutron EDM might start to probe the viable parameter space.

We also explored the region of parameter space with nonzero $\lambda_{22}^d$ instead of $\lambda_{32}^d$.  In this case, for BM1 and BM3, the neutron EDM is dominated by the strange quark contribution (\ref{eq:s-neturon-edm}), so its projected sensitivity covers larger region of parameter space.  However in this case, we did not find any viable region of parameter space explaining the anomaly in $a_\mu$.

%%%%%%%%%%%%%%%%%%%%%%%%%%%%%%%%%%%%%%%%%%%%%
\subsection{Probing the parameter space using electron measurements} \label{sec:electron}
%%%%%%%%%%%%%%%%%%%%%%%%%%%%%%%%%%%%%%%%%%%%%

Instead of muon specific couplings that address the discrepancies in the LFU ratios $R_{K^{(*)}}$ by new physics that suppresses the $b\to s\mu\mu$ transitions, one can also entertain the possibility that new physics addresses the anomaly by enhancing the $b\to see$ transitions. 
This can be achieved with the leptoquark couplings $\lambda^d_{31}$, $\lambda^d_{21}$ as given in Eq.~\eqref{eq:C9e} and by our benchmark points BM2 and BM4. 

These couplings will also lead to shifts in the anomalous magnetic moment of the electron, $\Delta a_e$, and, in the presence of CP violation, induce an electron EDM, $d_e$, (see  Eqs.~\eqref{eq:a_lepton} and ~\eqref{eq:d_lepton}, respectively), and the lepton flavor violating mode $\tau\to e\gamma$ (see Eq.~\eqref{eq:taumugamma} with $|\lambda_{32}^q \lambda_{33}^{d*}|^2\rightarrow|\lambda_{31}^d \lambda_{33}^{q*}|^2$).  Note that the chiral enhancement of the dipole moments $m_b/m_e$ can be particularly pronounced in the case of the electron. 

In this scenario, potentially important constraints arise from the $B_s \to e^+ e^-$ decay.
The effect of the leptoquark is given by an expression analogous to Eq.~\eqref{eq:Bsmumu} with $m_\mu \to m_e$ and $\lambda^f_{32}, \lambda^f_{22} \to \lambda^f_{31}, \lambda^f_{21}$, with the SM prediction given by BR$(B_s \to e^+e^-) = (8.54 \pm 0.55) \times 10^{-14}$~\cite{Bobeth:2013uxa}. 
Experimentally, the $B_s \to e^+e^-$ branching ratio is bounded at the 90\% C.L. by~\cite{Aaltonen:2009vr}
\begin{equation} \label{eq:Bsee_exp}
    \text{BR}(B_s \to e^+e^-) < 2.8 \times 10^{-7} ~.
\end{equation}

\begin{figure*}[t]
\centering
\includegraphics[width=0.45\textwidth]{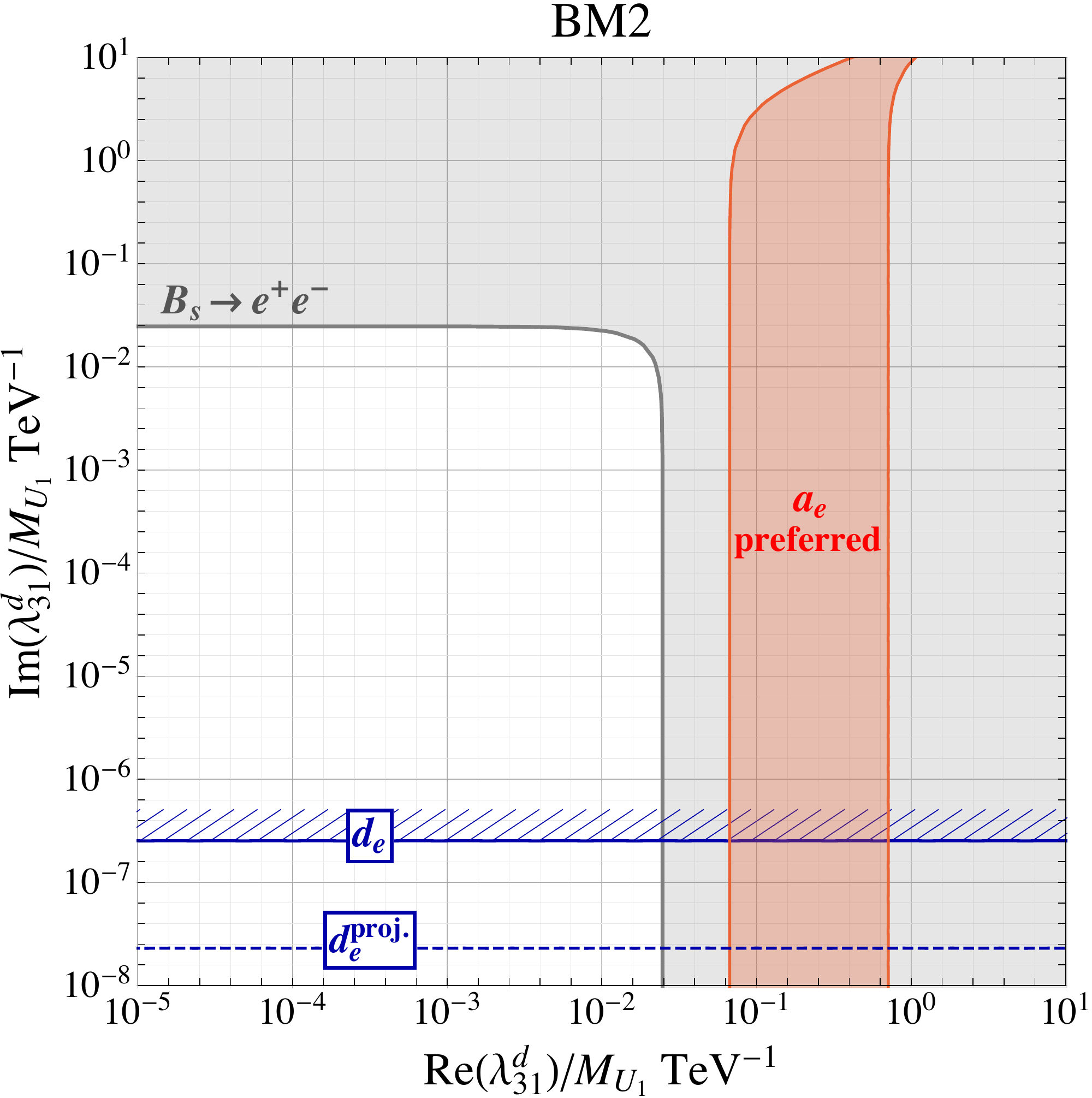} ~~~~
\includegraphics[width=0.45\textwidth]{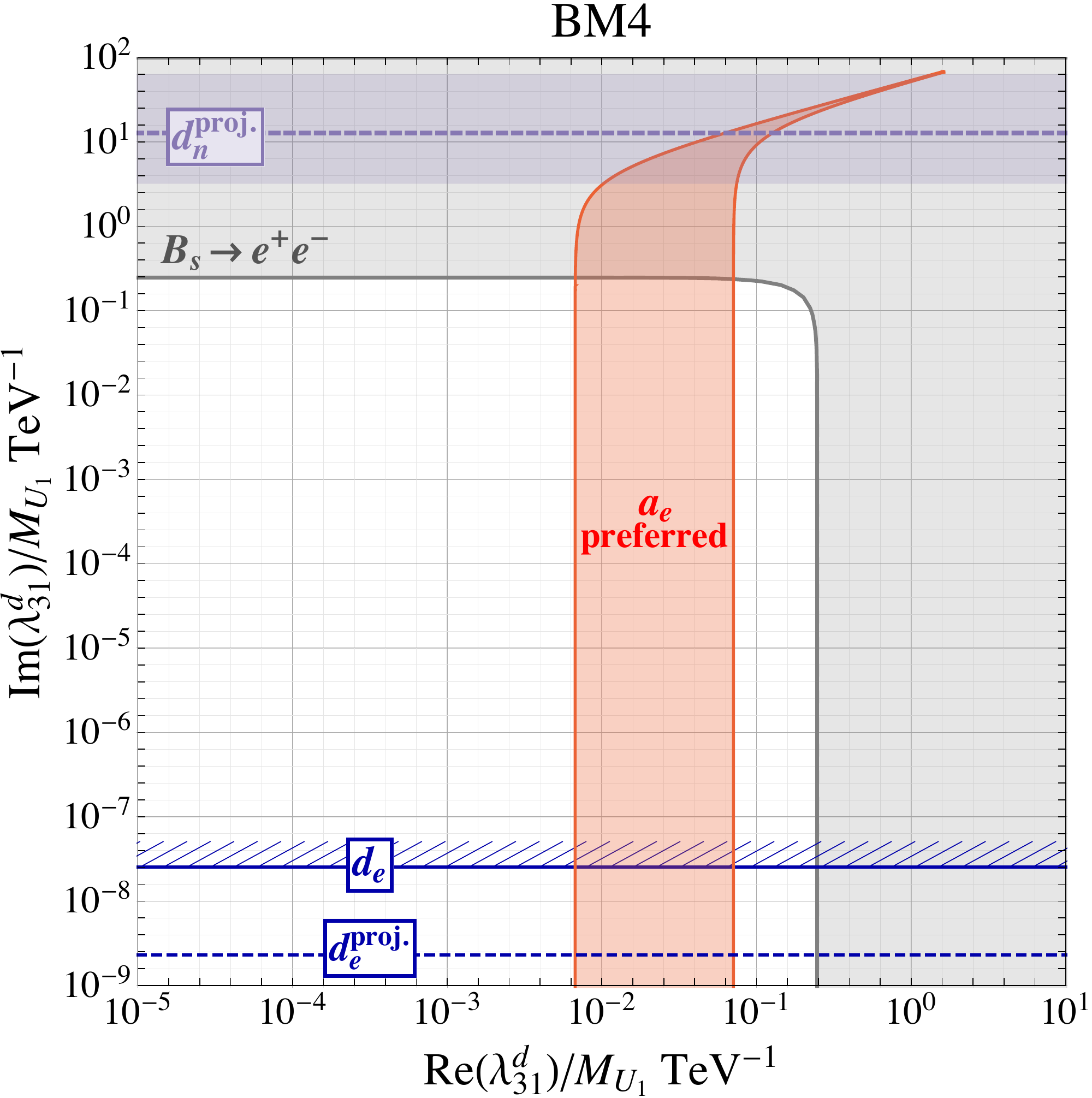}
\caption{Constraints on the $U_1$ leptoquark parameter space in the plane of the complex coupling $\lambda^d_{31}$ divided by the leptoquark mass for the benchmark points BM2 and BM4, left and right panel, respectively). The gray region is excluded by $B_s \to e^+e^-$ at the 95\% C.L.. The red shaded region corresponds to the parameter space the can address the anomaly in the anomalous magnetic moment of the electron. The solid (dashed) blue lines represent the current constraint (projected sensitivity) from the electron EDM. In the right panel, the dashed purple line represents the projected sensitivity from the neutron EDM, with the purple band reflecting the uncertainty in the nucleon matrix element $\beta^{\tilde{G}}_n$.}
\label{fig:electron}
\end{figure*}

The plots in Fig.~\ref{fig:electron} show the current and projected constraints on the $U_1$ leptoquark in the plane of the complex coupling $\lambda^d_{31}$ divided by the leptoquark mass for BM2 (left) and BM4 (right).  In both panels the gray region is excluded by the bound from $B_s \to e^+ e^-$, while the red shaded region is the region of parameter space that can address the $2.4 \sigma$ anomaly in the electron magnetic moment, $a_e$. The blue solid (dashed) lines are the current constraint (projected sensitivity) of the electron electric dipole moment, $d_e$. In the right panel, the dashed purple line and the surrounding purple band is the projected sensitivity of the neutron EDM, $d_n$.

For BM2 (left plot of Fig.~\ref{fig:electron}) we observe that the region of parameter space that is able to address the anomaly in $a_e$ is excluded by constraints from $B_s \to e^+ e^- $ and a simultaneous explanation of all the $B$ anomalies and $a_e$ is not possible. This is due to stringent constraints on the size of $\lambda^q_{31}$ from the lepton flavor violating decays $\tau \to \phi e$ and $B \to K \tau e$ (see discussion in Sec.~\ref{sec:LQBdecays}).  Constrains from the $\tau\to e\gamma$ are slightly weaker.

To avoid the stringent constraints from lepton flavor violating decays, we can set all the $U_1$ couplings to tau leptons to zero. Then, the $\tau \to \phi e$ and $B \to K \tau e$ rates as well as the $\tau \to e \gamma$ rate go to zero, and the left-handed couplings to electrons can be larger. However, by setting $\lambda^q_{33}$ and $\lambda^q_{23}$ to zero, we forgo an explanation of $R_{D^{(\ast)}}$. This scenario is given by BM4, and the resulting constraints are shown in the right plot of Fig.~\ref{fig:electron}. We observe that the smaller value of $\lambda^q_{21} = 0.005$ in BM4 leads to weaker constraints on $\lambda^d_{31}$ from $B_s \to e^+ e^-$. In addition, the larger value of $\lambda^q_{31} = 0.5$ generates a larger contribution to the electron magnetic moment necessary to explain the slight tension in $a_e$. In moving from BM2 to BM4 the bound from $B_s \to e^+e^-$ opens up a wide region in parameter space favorable for the electron magnetic moment, $a_e$. We conclude that BM4 can address the anomalies in both $R_{K^{(\ast)}}$ and $a_e$.

We also investigated the region of parameter space with nonzero $\lambda_{21}^d$ instead of $\lambda_{31}^d$.  We find in BM2 and BM4 that sensitivity to $d_e$ is reduced because it is chirally enhanced by $m_s$ rather than $m_b$ in Eq. (\ref{eq:d_lepton}).  We also find no region of parameter space where the $U_1$ leptoquark explains the tension of the measured $a_e$ with theory.

\section{LHC Bounds on the Leptoquark}\label{Sec:LHC}

{\renewcommand{\arraystretch}{1.1} 
\begin{table*}[tbp] 
\begin{ruledtabular}
\begin{tabular}{c c c }
\toprule
    \multicolumn{3}{c}{\textbf{LHC Bounds on Scalar Leptoquarks}} \\ \hline
   Channel & Experiment & Limit \\\hline
       \multicolumn{3}{c}{\textit{First Generation Leptoquarks}} \\ \hline
   \multirow{2}{*}{$eejj~(\beta =1 ) $} & ATLAS \cite{Aaboud:2019jcc} &  1400 GeV\\[-2.mm]
   & CMS \cite{Sirunyan:2018btu} & 1435 GeV \\   
  \multirow{2}{*}{$e\nu jj~(\beta =0.5)$}    &  ATLAS \cite{Aaboud:2019jcc} &  1290 GeV\\[-2.mm]
  &   CMS \cite{Sirunyan:2018btu} & 1270 GeV   \\
   \hline
      \multicolumn{3}{c}{\textit{Second Generation Leptoquarks}} \\ \hline
   \multirow{2}{*}{$\mu\mu jj ~(\beta = 1)$}&  ATLAS \cite{Aaboud:2019jcc}&1560 GeV\\[-2.0mm]
   & CMS \cite{Sirunyan:2018ryt} & 1530 GeV\\
  \multirow{2}{*}{$\mu\nu jj~(\beta = 0.5)$}  & ATLAS \cite{Aaboud:2019jcc}&1230 GeV \\[-2.0mm]
  &  CMS \cite{Sirunyan:2018ryt} & 1285 GeV \\
  \hline
   \multicolumn{3}{c}{\textit{Third Generation Leptoquarks}} \\ \hline
     \multirow{2}{*}{$b\tau b\tau$}  & ATLAS \cite{Aaboud:2019bye}& 1030 GeV \\
     &  CMS \cite{Sirunyan:2018vhk}& 1020 GeV\\  \hline
       \multicolumn{3}{c}{\textit{Reinterpreted SUSY searches}} \\ \hline
    $q\nu q\nu$&  CMS \cite{Sirunyan:2018kzh} &  980 GeV \\
     \multirow{2}{*}{$t\nu t\nu$} &  ATLAS \cite{Aaboud:2019bye} & 1000 GeV \\[-1.5mm]
     & CMS \cite{Sirunyan:2018kzh} &  1020 GeV  \\
     \hline \hline
     \multicolumn{3}{c}{\textbf{LHC Bounds on Vector Leptoquarks}} \\ \hline
Channel & Experiment & Limit \\\hline
\multicolumn{3}{c}{\textit{Reinterpreted SUSY searches}} \\ \hline
\multirow{2}{*}{$q\nu q\nu$}&  \multirow{2}{*}{CMS \cite{Sirunyan:2018kzh}} & 1410 GeV~($\kappa_s = 0$)\\[-2.0mm]
    && 1790 GeV~($\kappa_s = 1$)\\[2.5mm]
     \multirow{2}{*}{$t\nu t\nu$} 
     &  \multirow{2}{*}{CMS \cite{Sirunyan:2018kzh}} & 1460 GeV~($\kappa_s = 0$)\\[-2.0mm]
    && 1780 GeV~($\kappa_s = 1$)
\end{tabular}
\end{ruledtabular}
\caption{LHC bounds on pair-production of scalar and vector leptoquarks. For scalar leptoquarks, the first three sections correspond to bounds from dedicated leptoquark searches, while the last section corresponds to bounds derived from the reinterpretation of squark pair production searches. For vector leptoquarks, only reintepreted SUSY searches exist. The parameter $\beta$ denotes the branching ratio of the leptoquark to a quark and a charged lepton. We do not report the bounds on the decays of the LQ to down-type quarks and a neutrino since these decays do not exist in our model.}
\label{tab:LHCbounds}
 \end{table*}}
 
Low-energy flavor observables like those discussed in the previous sections provide an indirect probe of the $U_1$ leptoquark. A complementary approach to probe the existence of $U_1$ is direct production at high energy colliders and looking for signatures of their decay products. The goal of this section is to compute the lower bound on the leptoquark mass in the allowed regions of parameter space in Figs. \ref{fig:tau} -\ref{fig:electron}.

The two main production mechanisms are single production in association with a lepton ($gq \rightarrow \ell~U_1$), and pair production ($ g g, q \bar q \rightarrow U_1~\overline{U_1}$). For a recent review see \cite{Diaz:2017lit}. 
Once produced, the leptoquark will decay into a pair of SM fermions. The interactions of the $U_1$ leptoquark with SM quarks and leptons in Eq.~(\ref{eq:yuk}) generate the decays of $U_1$ into an up-type quark and a neutrino, or a down-type quark and a charged lepton. In the limit where $M_{U_1}$ is much larger than the masses of the decay products, the partial widths of $U_1$ are given by
\begin{subequations}\label{eq:decays}
\begin{align}
\Gamma(U_1 \rightarrow u_i \nu_j) &= \frac{M_{U_1}}{24 \pi} \bigg| \sum_{k = 1,2,3} V_{ik} \lambda^q_{kj} \bigg|^2\,, \\
\Gamma(U_1 \rightarrow d_i \ell_j) &= \frac{M_{U_1}}{24 \pi} \Big(  \big|\lambda^q_{ij}\big|^2 +  \big|\lambda^d_{ij}\big|^2 \Big)\,,
\end{align}
\end{subequations}
where $i,j = 1,2,3$ label the three generations.

Several dedicated searches for singly and pair produced scalar leptoquarks have been performed by the LHC, and are classified according to whether the leptoquark decays to first, second, or third generation fermions. The strongest bounds on leptoquark pair-production from ATLAS and CMS have been compiled in Tab. \ref{tab:LHCbounds}, where the searches are organized according to whether the branching ratio into a quark and a charged lepton (denoted by $\beta$) is 100\% or 50\%, with the remaining 50\% to a quark and a neutrino. In addition, in the table we also report the CMS reinterpretation of the squark pair production searches to place constraints on pair produced vector leptoquarks decaying to a quark and a neutrino, $t\nu$, or $q\nu~(q  = u,c,d,s)$  \cite{Sirunyan:2018kzh}. Similarly, ATLAS have presented reinterpretations of squark searches  \cite{Aaboud:2019bye}, although they only consider the decay of a leptoquark into 3rd generation quarks. We note that the ATLAS and CMS searches also consider leptoquark decays into down-type quarks and a neutrino (e.g $b \nu b \nu$ final states), but the corresponding couplings do not exist in our model and, therefore, we do not consider them here.

Singly produced scalar leptoquarks have been searched in $ej$, $\mu j$, and $b \tau$ final states.  The bounds on the leptoquark mass from single production depends on the coupling of the leptoquark to quarks and leptons. For unit couplings, 8 TeV searches for single production of first and second generation scalar leptoquarks constrain the leptoquark mass to be above $\sim 1700$ GeV and $\sim 700$ GeV, respectively \cite{Khachatryan:2015qda}, while the 13 TeV search for third generation scalar leptoquarks constrains the mass to be above 740 GeV \cite{Sirunyan:2018jdk}. In our benchmark models, the leptoquarks are mainly coupled to bottom or strange quarks. For this reason, the searches for singly produced %second and third generation 
leptoquarks are less sensitive to our benchmark models than the searches for pair produced leptoquarks. In the following, we will discuss in some details the bounds from searches of pair produced leptoquarks in all benchmarks.
%than the searches for pair produced leptoquarks and we do not consider them. On the other hand, the search for single production of first generation scalar leptoquarks will be relevant for the scenario in which the $U_1$ leptoquark couples dominantly to electrons in BM4.

For BM1 and BM2, the dominant non-zero couplings of $U_1$ are couplings involving tau leptons ($\lambda_{33}^q,\lambda_{23}^q$) and the dominant decay modes are $U_1 \to b \tau, s \tau, t \nu_\tau, c \nu_\tau$. At small values of $\lambda^d_{33}$ (see Fig. \ref{fig:tau}), the branching ratios of the $b\tau$ and $\tau \nu_\tau$ decay modes are similar in value ($\sim 0.25$) and dominate over the $s\tau$ and $c \nu_\tau$ decays modes, which themselves have similar branching ratios ($\sim 0.18$). For values of $\lambda^d_{33}$ near the border of the region allowed by $B_s \to \tau^+ \tau^-$ (see Fig. \ref{fig:tau}), the decay into $b\tau$ becomes the dominant decay mode with BR$(U_1\to b\tau)\sim 0.4$.

The reinterpreted SUSY search for pair production of vector leptoquarks decaying to $t \nu$ \cite{Sirunyan:2018kzh} and the CMS search for leptoquarks decaying to $b \tau$ \cite{Sirunyan:2018vhk} are the most sensitive searches. We find that these searches yield a similar lower bound on the mass of $U_1$ at around 1.2 TeV in the region of parameter space with small $\lambda^d_{33}$. The exact bound varies by at most $\sim$100 GeV in the region allowed by $B_s\to\tau^+ \tau^-$.

In BM3, $U_1$ couples dominantly to 2nd generation leptons and the main decay modes are $U_1 \to b\mu, s\mu, t \nu_\mu, c \nu_\mu$, with the $b \mu$ and $t \nu_\mu$ decays modes being the dominant ones since $\lambda^q_{32} \gg \lambda^q_{22}$, BR$(U_1\to t\nu_\mu)\sim{\rm{BR}}(U_1\to b\mu)\sim 0.5$. The most stringent LHC constraint on this benchmark comes from the search for pair produced leptoquarks in final states with two muons and two jets in  \cite{Sirunyan:2018ryt}\footnote{The search does not require any anti-$b$ tagging, and, therefore, we can simply apply it to our benchmark.}. This search leads to the bound $m_{U_1}\gtrsim 1.9$ TeV. This bound is valid in the entire parameter space shown in the right panel Fig. \ref{fig:muon}, since $\lambda^d_{32}$ is constrained to be very small, and therefore does not affect the leptoquark branching ratios.

Finally, in BM4, $U_1$ couples dominantly to 1st generation leptons and the main decay modes are $U_1 \to b e$ and $U_1 \to t \nu_e$.  In particular, at small values of $\lambda^d_{31}$ (see Fig. \ref{fig:electron}), the branching ratios of these decay modes are very similar in value ($\sim 0.5$). At larger values of $\lambda^d_{31}$, the branching ratio into $be$ becomes the dominant one, with BR$(U_1\to be)\sim 0.7$ at the border of the allowed region for $\lambda^d_{31}$, as shown in the right plot of Fig.~\ref{fig:electron}. The search for pair produced leptoquarks decaying in an electron and a jet in \cite{Sirunyan:2018btu} provides the strongest constraint on the mass of $U_1$ and gives a lower bound of $\sim 1.8$ TeV at small values of $\lambda^d_{31}$. The exact bound varies by at most $\sim$100 GeV in the region allowed by $B_s\to e^+ e^-$.

%The search for singly produced leptoquarks decaying to an electron and a jet places a similar bound on the mass of $U_1$ as the search for pair produced leptoquarks. Naively, the constraint from \cite{Khachatryan:2015qda} is expected to place a larger bound due to the larger production cross section of vector leptoquarks compared to scalar leptoquarks. However, the cross section must be scaled appropriately by the branching BR$(U_1 \to be)$ and the square of the coupling, which counteracts the enhancement in the production cross section. This leads to a bound on the  mass of $U_1$ that is comparable to that from pair produced leptoquarks in the region allowed by $B_s \to e^+ e^-$.

%%%%%%%%%%%%%%%%%%%%%%%%%%%%%%%%%%%%%%%%%%%%
\section{Conclusions}\label{Sec:Conclusions}
%%%%%%%%%%%%%%%%%%%%%%%%%%%%%%%%%%%%%%%%%%%%
In this study, we focused on the possible, and quite likely, existence of new sources of CP violation if the flavor anomalies in $b \to c$ and $b \to s$ decays are due to new physics,  specifically in the case where the new physics consists of a $U_1$ vector leptoquark. The underpinning of our study is that the $U_1$ vector leptoquark is one of the only (if not {\em the} only) new physics scenarios known to us that can provide a simultaneous explanation of the anomalies observed in lepton flavor universality ratios in $b \to c \ell \nu$ and $b \to s \ell \ell$ decays, $R_{D^{(*)}}$ and $R_{K^{(*)}}$.
Since the couplings of the $U_1$ to quarks and leptons are generically CP violating, they are expected just as generically to produce potentially observable electric dipole moments (EDMs) in leptonic and hadronic systems. Here, we have first provided new, original, and complete formulae for the calculation of the relevant EDMs, and carried out a phenomenological study of a few benchmark cases of how EDMs can constrain the $U_1$ leptoquark interpretation of the anomalies.

We note that the expressions we provided are the most general expressions for dipole moments induced by vector leptoquarks at one loop level, accounting for the most generic set of leptoquark couplings, which can accomodate scenarios for which the leptoquark may be composite.  

We explored the parameter space of the $U_1$ leptoquark in the vicinity of 4 benchmark points that explain the $R_{D^{(*)}}$ and $R_{K^{(*)}}$ anomalies (or a subset of them). 
We identified viable regions of parameter space where the existing discrepancies in the anomalous magnetic dipole moments of the electron $a_e$ and the muon $a_\mu$ can be explained in addition to $R_K^{(*)}$.
However, we concluded that a simultaneous explanation of all three classes of discrepancies ($R_{D^{(*)}}$, $R_{K^{(*)}}$, $a_{e,\mu}$) is not possible.

We found that, in the presence of non-zero CP-violating phases in the leptoquark couplings, EDMs play an important role in probing the parameter space of the model.  Existing bounds on the electron EDM already exclude large parts of parameter space with CP violating leptoquark couplings to electrons.  The expected sensitivities to the neutron EDM can probe into motivated parameter space and probe imaginary parts of leptoquark couplings to taus and muons.

%%%%%%%%%%%%%%%%%%%%%%%%%%%%%%%%%%%%%%%%%%%%%%%
\section*{Acknowledgements}
%%%%%%%%%%%%%%%%%%%%%%%%%%%%%%%%%%%%%%%%%%%%%%%
The research of WA is supported by the National Science Foundation under Grant No. PHY-1912719. WA acknowledges support by the Munich Institute for Astro- and Particle Physics (MIAPP) which is funded by the Deutsche Forschungsgemeinschaft (DFG, German Research Foundation) under Germany's Excellence Strategy - EXC-2094 - 390783311.
The research of SG and DT is supported in part by the NSF CAREER grant PHY-1915852. SG would like to thank the Aspen Center for Physics under NSF grant PHY-1607611, where part of this work was performed.
HHP and SP are partly supported by the U.S.\ Department of Energy grant number de-sc0010107.
The work of DT is supported in part by the U.S. Department of Energy, Office of Science, Office of Workforce Development for Teachers and Scientists, Office of Science Graduate Student Research (SCGSR) program. The SCGSR program is administered by the Oak Ridge Institute for Science and Education for the DOE under contract number de-sc0014664. 

%%%%%%%%%%%%%%%%%%%%%%%%%%%%%%%%%%%%%%%%%%%%
\bibliographystyle{atlasBibStyleWithTitle.bst}
\bibliography{bib}

\end{document}